%% file: icrawl_jcdl15.tex
\documentclass{sig-alternate-2013}
\pdfoutput=1 
\usepackage{balance}
\usepackage[
  pass,
]{geometry}

\newtheorem{mydef}{Definition}[section]
\usepackage{subfig}
\usepackage{tabularx}

\usepackage[T1]{fontenc}
\usepackage[utf8]{inputenc}
\usepackage{booktabs}
\usepackage{xspace}
\usepackage[square,numbers]{natbib}
\usepackage{graphicx}
\usepackage{color}
\usepackage{hyperref}

\newcommand{\eg}{e.g.\@\xspace}
\newcommand{\UN}{UN\xspace}
\newcommand{\FO}{FO\xspace}
\newcommand{\INT}{INT\xspace}
\newcommand{\TB}{TB\xspace}
\newcommand{\Ebola}{Ebola Epidemic\xspace}
\newcommand{\Ukraine}{Ukraine Tensions\xspace}

\newfont{\mycrnotice}{ptmr8t at 7pt}
\newfont{\myconfname}{ptmri8t at 7pt}
\let\crnotice\mycrnotice%
\let\confname\myconfname%

\permission{Permission to make digital or hard copies of all or part of this work for personal or classroom use is granted without fee provided that copies are not made or distributed for profit or commercial advantage and that copies bear this notice and the full citation on the first page. Copyrights for components of this work owned by others than ACM must be honored. Abstracting with credit is permitted. To copy otherwise, or republish, to post on servers or to redistribute to lists, requires prior specific permission and/or a fee. Request permissions from permissions@acm.org.}
\conferenceinfo{JCDL'15,}{June 21--25, 2015, Knoxville, Tennessee, USA.\\
{\mycrnotice{Copyright is held by the owner/author(s). Publication rights licensed to ACM.}}}
\copyrightetc{ACM \the\acmcopyr}
\crdata{978-1-4503-3594-2/15/06\ ...\$15.00.\\
DOI: http://dx.doi.org/10.1145/2756406.2756925}

\begin{document}
%
\conferenceinfo{JCDL}{2015}

%
%
\title{iCrawl: Improving the Freshness of Web Collections by Integrating Social Web and Focused Web Crawling}

\numberofauthors{1} 
\author{
\alignauthor
Gerhard Gossen, Elena Demidova, Thomas Risse\\
       \affaddr{L3S Research Center, Hannover, Germany}\\
       \email{\{gossen, demidova, risse\}@L3S.de}
}

\maketitle
\begin{abstract}
Researchers in the Digital Humanities and journalists need to monitor, collect and analyze fresh online content regarding current 
events such as the Ebola outbreak or the Ukraine crisis on demand.  
However, existing focused crawling approaches 
only consider topical aspects while ignoring temporal aspects and therefore cannot achieve thematically coherent and fresh Web collections.
Especially Social Media provide a rich source of fresh content, which is not used by  state-of-the-art focused crawlers.
In this paper we address the issues of enabling the collection of fresh and relevant Web and Social Web content for a topic of 
interest through seamless integration of Web and Social Media in a novel integrated focused crawler. 
The crawler collects Web and Social Media content in a single system and exploits the stream of fresh 
Social Media content for guiding the crawler.
\end{abstract}

%
%

\section{Introduction}
\label{sec:intro}
\input{input/introduction}



\input{input/integrated-focused-crawling}

\input{input/architecture}



\input{input/evaluation}

\section{Related Work}
\label{sec:related-work}
\input{input/related}

\section{Conclusion}
\label{sec:conclusion}
\input{input/conclusion}
\balance

\small
\bibliographystyle{abbrvnat}
\bibliography{dl}

\end{document}

%% file: input/introduction.tex

With the advancement of analysis and mining technologies, a growing interest in collecting and analyzing Web content can be observed in various scientific disciplines. Our user requirements study~\cite{icrawlRequirements} revealed that more and more disciplines are interested in mining and analyzing the Web. User-generated content and especially the Social Web is attractive for many humanities disciplines. Journalists are also interested in the content as it provides a direct access to the people's views about politics, events, persons and popular topics shared on Social Media platforms. These users require a comprehensive on-demand documentation of the activities on the Web around global events (e.g. Ebola outbreak, Ukraine crisis) and local events (e.g. Blockupy, squatting). A comprehensive documentation consists of official communications, news articles, blogs, and Social Media content. 

Both the Web and Social Media can provide a wealth of information and opinions about emerging events and topics. Often these media are used complementary, in that discussions about documents on the Web occur on Social Media or that Social Media users publish longer posts as articles on Web sites. Organizations generating fresh Web content, e.g. news agencies, often offer entry points to this content via Social Media such as Twitter. These are taken up by other Twitter and Facebook users for recommendation and discussion with other users. During the discourse further links are recommended. Especially Twitter turns out to be one of the most popular media to spread new information. 

Large scale analysis and mining of Web and Social Media content requires that the relevant content is stored in easily accessible collections (as opposed to being distributed across the Web). The user requirements posed on such collections include most importantly topical relevance, freshness, and context \cite{icrawlRequirements}. As the Web and Social Web are ephemeral and under constant evolution, Web pages can quickly change or become unreachable even within hours. Content linked from Social Media content typically has a very short life span \citep{salaheldeen2012}, which increases the risk of missing Web content linked from Social Media unless it is collected immediately. Timely collection of the embedded information (like embedded Twitter feeds) plays a crucial role in building comprehensive collections that cover all important media on the Web. On the other hand, tweets containing embedded links should be accompanied by the linked pages to understand their context. 

In order to create Web collections on demand, Web crawlers are gaining interest in the community. Web crawlers are automatic programs that follow the links in the Web graph to gather documents.  Unfocused Web crawlers (e.g. Heritrix \cite{mohr04heritrix} and Apache Nutch \cite{nutch}) are typically used to create collections for Web archives or Web search engines, respectively. These crawlers collect all documents on their way through the Web graph and produce vast document collections on a variety of topics. In contrast, focused Web crawlers \citep[e.g.~][]{Aggarwal:2001:ICW:371920.371955,chakrabarti99:focus} take topical or temporal \cite{PereiraMCM14} dimensions of the collected pages into consideration. 

\begin{figure*}[th!]
\centering
\includegraphics[width=1.0\textwidth]{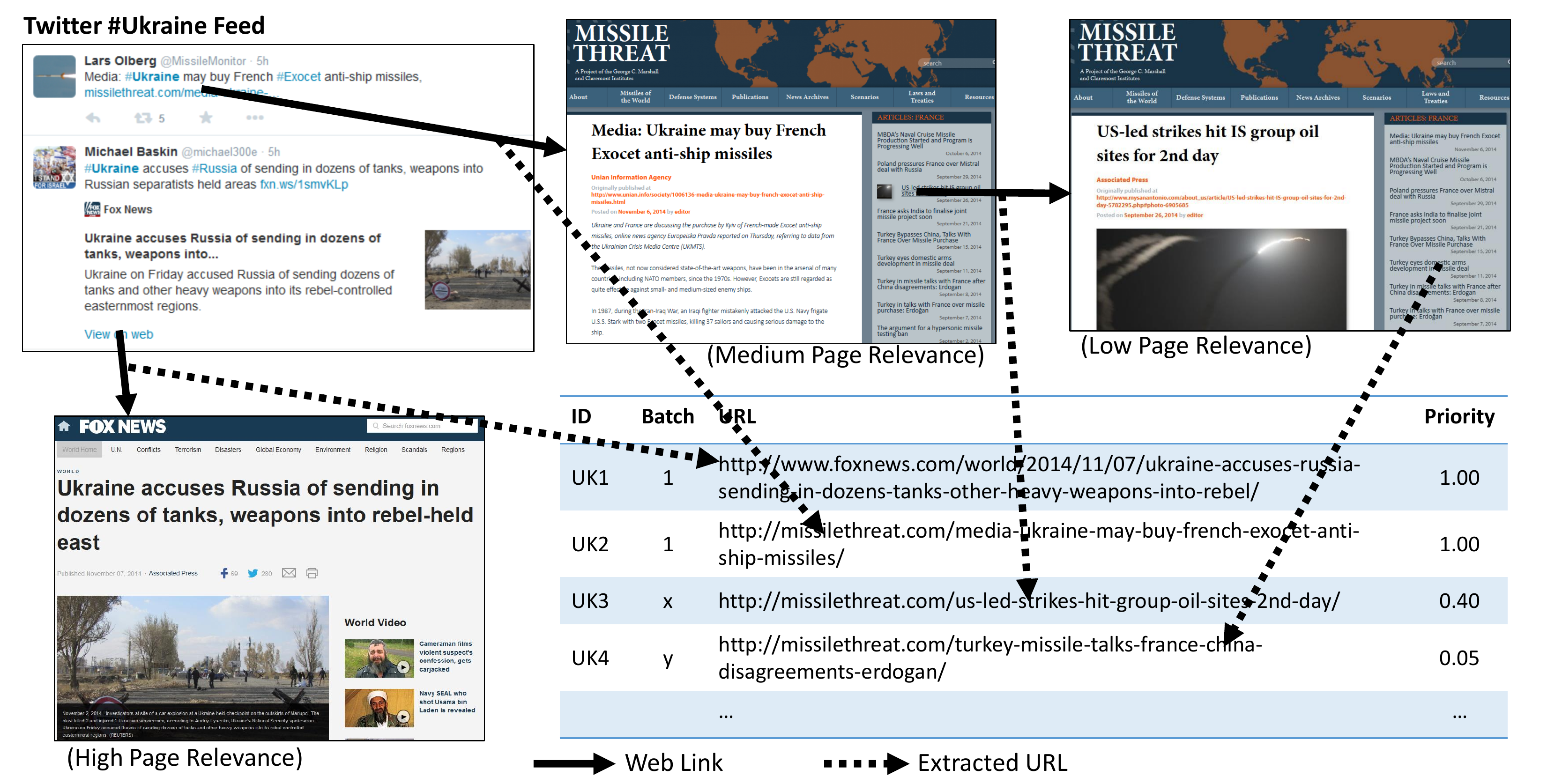}
\caption{Interlinking of Twitter and Web and resulting Crawler Queue}
\label{fig:interlink}
\end{figure*}

Existing focused crawling approaches consider topical and temporal aspects in isolation, and thus failing to create collections of Web documents that are not only thematically coherent, but also up-to-date. Furthermore, most crawlers are dedicated to collect either Web \citep[e.g.~][]{Aggarwal:2001:ICW:371920.371955,chakrabarti99:focus}  or Social Web content \citep[e.g.~][]{Boanjak:2012:TDF:2187980.2188266, psallidas2013}. Even in crawlers that make use of both Web and Social Web content like the ARCOMEM crawler \cite{RisseFI:2014}, the connection is rather loose as Social Web content is collected independently of Web content. The usage of loosely coupled crawlers can easily lead to a big time gap between the collection of the content and linked or embedded content. 

By reducing this time gap the freshness of the content can greatly be improved. Finding recent and relevant Web pages automatically is a difficult task for the Web crawler. In order to initiate the crawling process, existing Web crawlers require seed URLs, i.e. a set of pages to start the crawling. Typically, these seed URLs are provided by the researcher manually, such that crawl setup requires expert knowledge and lacks flexibility. Especially as recently created Web content may not (yet) be well interlinked on the Web, static crawl specifications can miss important entry points to collect fresh content appearing on the Web during the crawl. Therefore, it is important to provide means to automatically identify relevant and fresh content in a continuous way \cite{Yang:2012}.

In contrast to Web crawling, Social Media sites like Twitter or Flickr provide access to their selected content through custom APIs. For example, the Twitter streaming API allows the collection of fresh and topically relevant tweets using standing queries making it possible to focus the requested stream on the topic of interest.

In this paper we address the problem of enabling on demand snapshot collections of fresh and relevant Web content for a topic of interest. To this goal we combine the advantages of focused Web crawling and Social Media API queries in the novel paradigm of integrated crawling. In this paradigm we use recent tweets to continuously guide a focused crawler towards fresh Web documents on the topic of interest and to jointly collect recent and relevant documents from different sources.

The contributions of this paper are as follows: 
\vspace{-\topsep}
\begin{itemize}
\itemsep-2pt
  
\item We present the novel paradigm of integrated crawling that enables the continuous guidance of a focused crawler towards fresh and relevant content. In order to directly provide the Web crawler with entry points to fresh Web content, we exploit the idea of integrating Social Media with focused Web crawling. 

 \item We present iCrawl\footnote{Available at \url{http://icrawl.l3s.uni-hannover.de}. The system is unrelated to other similarly named projects.},  an open source 
 integrated crawler to perform focused crawls on  current events and topics on demand. The extensible iCrawl 
 architecture seamlessly integrates Twitter API query and Web crawling. To achieve scalability, our architecture 
 extends the Apache Nutch crawler \cite{nutch}. 

\item We demonstrate the efficiency and effectiveness of our approach in a number of experiments 
with real-world datasets collected by iCrawl. Specifically, we create and analyze collections about the Ebola epidemic and the Ukraine crisis.

\end{itemize}


%% file: input/integrated-focused-crawling.tex
\section{Integrating Web Crawling and Social Media}
\label{sec:integrated-focused-crawling}

Whereas Web content is typically collected through Web crawlers that follow outgoing links from Web pages, 
Social Media platforms such as Twitter supply fresh content via streaming APIs. Combining the two paradigms would 
allow us to follow recent events and discussions in both media in a seamless way. 

In the example shown in Figure~\ref{fig:interlink}, a researcher or a journalist is interested in creating a 
specific Web snapshot regarding reports that Russia was sending tanks to the East Ukraine on November 7, 2014. 
Fresh information around this event can be received from Twitter. Since it is early in the event and the impact 
is unclear, no specific hashtag exists. Therefore the more generic topic of \#Ukraine can be followed. 
From the \#Ukraine Twitter stream many links can be extracted around the Ukraine crisis, but not all of them 
are related to the topic of Russian tanks. For example, the posted link to the Fox News article is of high 
relevance as it describes the event itself. The relevance of the posted link from Missile Threat is on a 
medium level since it talks about Ukraine and missiles, but not about this specific event. 
Furthermore links from this unrelated page point to other unrelated pages, which have overall 
low relevance as they only talk about weapons in some form. The task of the crawler is to prioritize the extracted 
links for fetching according to their relevance as we discuss in Section~\ref{sec:prior-focus-freshn}.

Since the aim of our integrated crawler is the creation of collections with fresh and relevant content we 
first need to define freshness in the context of Web collections. Afterwards we present the 
architectural challenges for an integration of focused Web crawling with Social 
Media streams.

\subsection{Estimating Freshness of Web Content}
\label{sec:freshness}

Along with the topical relevance, freshness of the collected pages is one of the crucial requirements of 
journalists and researchers, who are interested in irregular crawls on current events and topics performed on demand. 
In this context, it is important to automatically guide the crawler towards relevant and fresh content and to estimate 
freshness of the crawled pages efficiently and accurately.
Intuitively, freshness of a page can be determined using the length of the time interval 
between the fetch time of this page and the (estimated) creation time. 

\begin{mydef}
\label{def:freshness}
\textbf{Page Freshness:} If the page $P$ was fetched at time $t_{f}$ and the (estimated) time of the creation 
of this page is $t_{c}$, then the freshness $F_{P}$ of $P$ is proportional to the length of the time interval 
between the fetch time and the (estimated) creation time: $F_{P} \approx t_{f} - t_{c}$.
\end{mydef} 

The creation time of collected pages can be easily estimated if a crawl history is available. This history is typically 
obtained by regular unfocused large-scale crawls and can show the changes in the Web graph as well as provide 
the date the document has first been discovered by the crawler (\emph{inception date}) \cite{dong2011Google}. 
This approach is used by many of the larger organizations in the field of Web crawling, e.g. Web search engines 
and Web archiving institutions.
In case of irregular on-demand crawls on current events and topics such crawl history may not be available. 
In this case the crawler can rely on the content-based creation date estimates.

Content-based estimation of Web page creation time faces technical challenges as such information is highly 
syntactically heterogeneous and not always explicitly available. Moreover, Web pages do not offer reliable 
metadata concerning their creation date \cite{Tannier14}. In order to allow an efficient freshness evaluation, 
in this paper we combine several features including page metadata and page content. Other possibilities include 
the usage of tools such as DCTFinder \cite{Tannier14} that combine rule-based approaches with probabilistic 
models and achieve high precision.

\subsection{Crawler Architecture}
\label{sec:crawler-architecture}
Web and Social Media crawling follow two different paradigms and are therefore handled separately in current systems.
Standard Web crawling is a \emph{pull} process: The crawler fetches pending URLs from its queue while taking into 
account the expected utility of the URLs, politeness requirements and other factors. In turn, the outlinks from the 
fetched documents are added to the queue to continue the crawling process. These steps are continuously repeated 
until the crawler is stopped. The pull characteristic of the process enables the crawler to control its strategy 
during the entire process, for example with regard to the crawl rate and the scope of the created collection.

In contrast, Social Media streaming APIs are based on a \emph{push} mechanism meaning that the crawler has to submit a 
fixed standing query to the API. 
The platform returns fresh content matching the query asynchronously.
%
This access mechanism does not give the crawler sufficient control over the input, as the rate and time intervals of 
the input from the Social Media API are not known in advance.

These differences together with the dynamic nature of Social Media platforms present several major challenges 
with respect to the seamless integration of the Web and Social Media crawling. 
Due to these differences, Social Media access cannot occur as part of the Web crawler loop, 
but has to be handled separately. However, the results of both processes are highly interdependent and 
impact each other in several ways. 
First, Social Media is a valuable source of relevant links to the Web (as well as to related streams in the Social Web).
Therefore, the filtered stream of relevant Social Media messages containing outgoing Web links 
need to be placed into the Web crawler queue. 
Second, one flavor of Social Media integration is the embedding of message feeds into Web pages. Since it can be 
assumed that the embedded feed is relevant to the page content, the feed content should also be 
collected. Therefore, the Web crawler needs to communicate such embedded feeds to the Social Media crawler so that it can collect relevant posts  by adjusting its queries.
Third, the highly dynamic nature of platforms like Twitter requires that the interaction between Web and Social Media crawlers is efficient and has a low latency to ensure that content can be captured effectively.

\subsection{Prioritization for Focus and Freshness}
\label{sec:prior-focus-freshn}

The Web crawler needs to decide at each step, which of the queued URLs it will crawl next. This is typically based on a \textbf{relevance score} for the URL 
based, for example, on the relevance of pages linking to that page, the importance of the web site host or (if the URL has been crawled before) the estimated change probability. Focused crawlers primarily use the content of the linking pages to estimate the relevance of the URL to the crawl focus, based on the assumption that Web pages typically link to other pages on the same topic.

In contrast, Social Media often only provide a limited amount of content that can be used for the relevance assessment. For example, Twitter only allows 140 characters for each post. However, Social Media platforms offer more expressive queries. For example, it is possible to retrieve posts matching certain keywords, posts written by specified users or posts from a given geographical region. This can allow the crawler to directly query the posts relevant to the crawl focus. Additionally, the crawler can make use of the Social Media specific measures of relevance, such as the popularity of a post or the profile of the posting user.


Existing focused Web crawlers typically aim to collect topically coherent Web content for a topic of interest \cite{chakrabarti99:focus}. Only few works have considered time as the focusing target \cite{PereiraMCM14}. Neither of these works have however considered the aspect of \textbf{freshness} (see Section~\ref{sec:freshness}) of the collected documents in the context of crawler prioritization. As we will show in this work, the use of Social Media makes it possible to create fresh and focused Web document collections.


%% file: input/architecture.tex
\section{Crawler Architecture}
\label{sec:architecture}

\begin{figure}
  \centering
  \includegraphics[width=\linewidth]{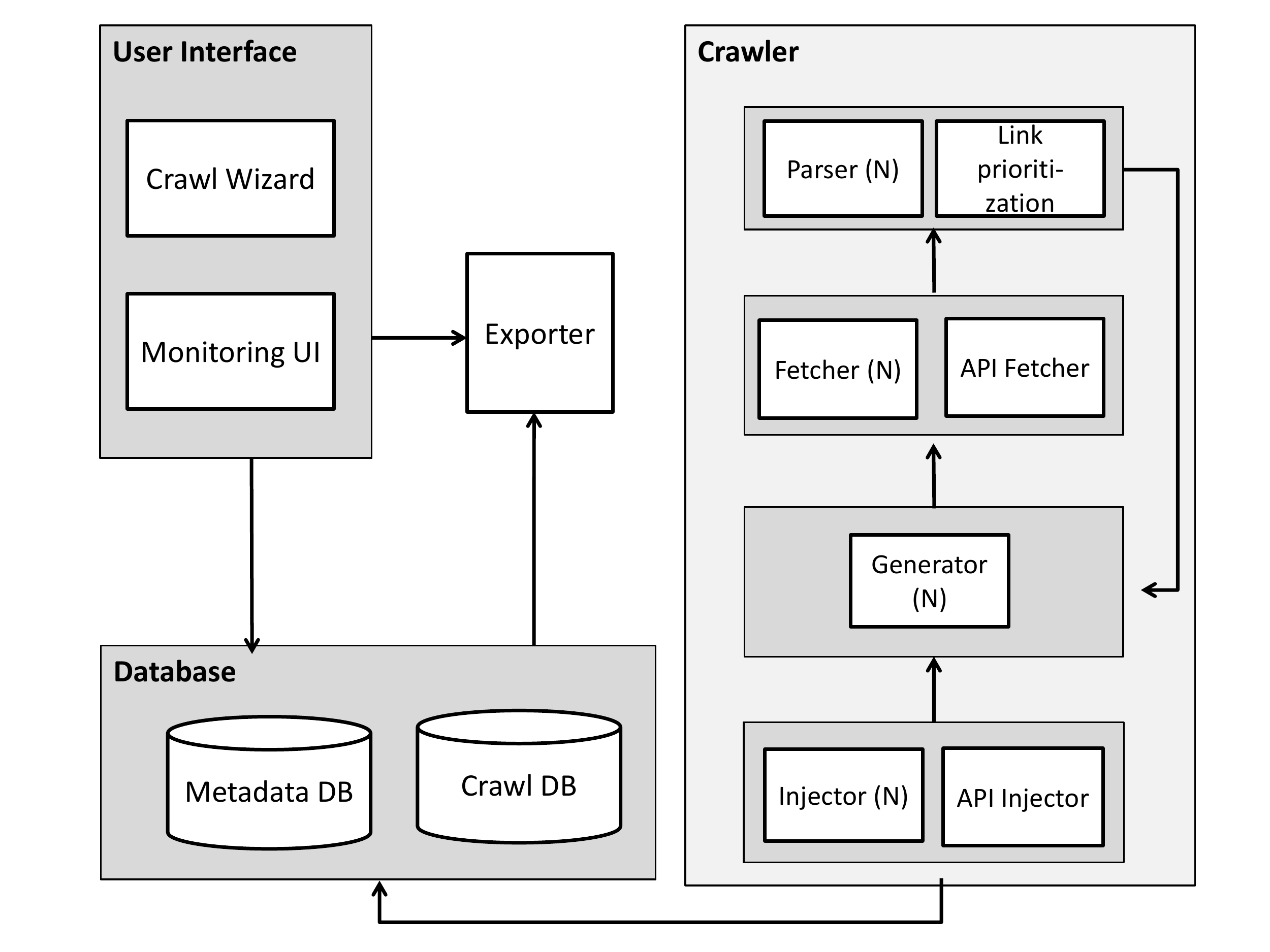}
  \caption{Architecture of the iCrawl system. Components marked with (N) are provided by Apache Nutch.}
  \label{fig:architecture}
\end{figure}
The iCrawl architecture (see Figure~\ref{fig:architecture}) is based on Apache Nutch~\cite{nutch},
an established Web crawler that is designed to run as a distributed system on 
the Hadoop map/reduce platform. Nutch provides the general functionality of a Web crawler, 
such as crawl queue management, fetching and link extraction. 
Nutch is highly extensible through a plugin system that provides extension points for 
all phases of the crawl. As Nutch is run as a series of map/reduce jobs, it can also 
be customized by replacing or inserting custom jobs into the crawling cycle. In order 
to implement an integrated and focused crawler, we modified the Nutch crawler 
and complemented it with additional modules.
Additionally, we created several components such as a graphical user interface that make the system more useful for our target users.
In the following we will describe the facets of our system relevant to integrated and focused crawling.

\subsection{Web Crawler}
\label{sec:web-crawler}

Nutch implements the basic Web crawler through collaborating map/reduce jobs (see Figure~\ref{fig:architecture}).
First, the seed URLs of a crawl are added to the queue by the \emph{Injector} job.
Then the crawl is executed as a series of batches.
Each batch starts with the \emph{Generator} job that picks $n$ URLs from the queue that have the 
highest priority and can be fetched right now.
The \emph{Fetcher} job downloads the Web pages for these URLs and stores them locally.
The \emph{Parse} 
job extracts links and metadata from the pages, calculates 
the priority for the outlinks and updates the crawler queue with the new values.
The \emph{Generator}, \emph{Fetcher} and \emph{Parse} jobs
are repeated until the crawl is stopped.

\subsection{Integrated Crawling}
\label{sec:integrated-crawling}

As described above, Nutch only adds seed URLs to the queue at the beginning of the crawl through the \emph{Injector} job.
However, Social media streams provide us constantly with new potential seed URLs.
We therefore implemented custom \emph{API Injectors} for URLs from Social Media streams to enable integrated crawling.
Currently, we provide support for the Twitter streaming API and RSS feeds; other sources can be added.
The crawler user can specify for each API the queries they want to monitor when starting the crawl.
Furthermore, additional queries can be added during the crawl manually or automatically and the current queries can be modified to reflect a shifting topic.

The API Injectors cannot run inside Nutch, as each Nutch job runs only for a short amount of time, whereas the \emph{push} nature of the APIs 
(see Section~\ref{sec:crawler-architecture}) requires a continuously running process.
Our system automatically starts the API Injectors for the requested sources and also shuts them down when the crawl is stopped.
When an API Injector is started, it receives the specified queries and is then responsible for creating and maintaining the API connection.
The API Injector inject URLs into the crawler queue, store received posts and add information about resolved redirects (e.g. from URL shorteners) through a simple interface modeled on Hadoop\footnote{The library to communicate with Nutch is available separately as open source at \url{https://github.com/L3S/nutch-injector}}.

An URL is only added to the crawler queue when it is first discovered.
If an URL was already discovered by crawler, its relevance is unchanged as the content in social media posts is typically to short to estimate relevance.

Social Media APIs are also used during the regular crawl to augment the Web crawler.
Through the APIs information about for example posts is available in formats such as JSON.
These are easier to process automatically and sometimes even contain additional information.
When we encounter links to known Social Media websites, we rewrite the links to point directly to the APIs and enqueue those links in addition to the Web URLs.
The crawler then calls the \emph{API Fetcher} module for the appropriate Social Media site when the URL needs to be fetched to retrieve the document through the API and store it in the crawl database.

The described process is illustrated in Figure~\ref{fig:interlink} where the Twitter stream is 
filtered for ``\#Ukraine''. The extracted links from the filtered stream 
(Fox News [UK1] and Missile Threat [UK2]) are added to the queue with a high priority 
of 1 and crawled in the first batch. 
After UK2 is crawled, its outgoing links are analyzed. This analysis results in a relevance 
score of 0.4 for the extracted link UK3 to be crawled in a later batch $x$. 
After the crawling of UK3 the analysis results in a low relevance of 0.05 for the outlink 
UK4 to be crawled at a later point in time (if at all).

The combination of Web crawler, API Injectors and API Fetchers allows us to seamlessly cross the boundaries between Web and Social Media in a single system.

\subsection{Focused Crawling}
\label{sec:focused-crawling}

As our goal is to create a topical Web collection, we need to ensure that only Web pages relevant to that topic are crawled.
However, Nutch only uses a relevance measure based on the link graph (Adaptive On-line Page Importance Computation~\cite{Abiteboul2003}).
This measure does not take the topic into account at all.
Furthermore, it requires multiple crawls of the same pages until the relevance predictions converge.
Therefore we replace the priority computation of Nutch with our own module (\emph{Link prioritization}).
It implements the prioritization by determining the relevance of each downloaded page to the crawl topic and computing a priority score for each of its outlinks.
These scores are returned to Nutch, which in a separate step combines them with scores from other pages and updates the crawler queue accordingly.
In this way URLs linked to from pages of high topical relevance are moved to the front of the queue, especially if they are linked to repeatedly, whereas the outlinks of low relevance pages are moved to the back of the queue.
When the relevance of pages at the front of the queue sinks below a threshold, we stop the crawl and ensure in this way that the collected pages are of overall high relevance.

\subsection{Data Storage}
\label{sec:data-storage}

During the \emph{Parse} job we extract and store entities and keywords from the crawled documents.
This metadata is used to provide an improved monitoring interface for the crawl and can be used for semantic indexing of the crawled data.
We also collect extensive metrics about the crawl itself to provide a good documentation.
This means that our crawl generates two different types of data with varying characteristics and access patterns, namely crawled content and metadata.
The crawled content is typically larger (several kilobytes or megabytes per record) and is frequently accessed in sequence, e.g. by analysis or export processes. 
This data is stored in the \emph{Crawl DB} backed by the distributed Apache HBase datastore which can easily store gigabytes or terabytes in a fault-tolerant manner.
On the other hand, metadata is smaller in size (less than a kilobyte per record), but needs to be accessed and queried in many different ways.
Standard relational databases work better for this data than HBase, therefore we store it in a separate \emph{metadata DB}.
By having these two data stores we can ensure a good performance of all components of our platform.

\subsection{WARC Exporter}
\label{sec:warc-exporter}

There are already many systems to provide index and analyze Web collection.
Rather than duplicate this effort, we provide a way to export the final collection in the standard WARC format. 
The exported files also contain the extracted metadata. 
This metadata can be used for exploration of the collection or can be indexed to provide richer search interfaces for the content.

\subsection{Crawl Specification and User Interface}
\label{sec:user-interface}

The crawling process starts with the manual definition of the \emph{crawl specification}:
a list of seeds (URLs and Social Media queries) and keywords that best 
describe the topical focus of the crawl from the user's point of view. 
The \emph{crawl specification} is used in two ways: 
(1) to support the focusing and prioritization of the crawl 
and (2) to provide an initial seed list for the crawler.
%
%
The setup and scoping of a crawl is supported through a `Wizard' interface that allows the user to find 
and select relevant Web and Social Media seeds through simple keyword queries. 
More details about the user interface can be found in \cite{gerhardgossen2015icrawl}.



%% file: input/evaluation.tex
\section{Evaluation}
\label{sec:evaluation}

The goal of the evaluation is to measure the impact of the Social Media integration on 
the freshness and the relevance of the resulting collections with respect to the 
topical focus of the crawl as well as to better understand 
the domains of the content covered using the different crawler configurations. 
To achieve this goal, in our evaluation we compare several crawler configurations that vary with respect 
to the focusing and integration: 

\begin{description}
\item[Unfocused (\UN):] Our first baseline is a typical unfocused state-of-the-art Web crawler.
For this configuration we use an unmodified version of Apache Nutch.
We expect this configuration to collect less relevant documents than the other configurations.
\item[Focused (\FO):] As a second baseline we incorporate state-of-the-art focusing features into Apache Nutch 
to get a focused crawler.
This configuration is expected to find more relevant documents than the unfocused crawler, 
but still does not take their freshness into account.
\item[Twitter-based (\TB):] To better understand the role of Social Media API input in the integrated crawler, 
we use a simple crawler that monitors the Twitter streaming API for a given query and downloads 
all documents linked from the returned tweets (without following further outlinks of those pages).
We expect that the tweeted links are typically very fresh.
\item[Integrated (\INT):] This configuration uses our proposed system and combines the focused Web 
crawler and the Twitter API input as described in Section~\ref{sec:architecture}. This configuration 
combines the advantages of the focused crawler and the Twitter API and is expected to deliver 
fresh and relevant results.   
\end{description}


For each of these configurations, we measure the relevance and freshness of the collected documents during the 
runtime of the crawl. An ideal system would have a constantly good relevance and freshness until all 
relevant documents have been crawled, after which the relevance has to drop.
However, in contrast to previous work on focused crawlers we target ongoing events, where new relevant 
documents can be created during the crawl.
This means that the crawler can have a continuous supply 
of relevant and fresh documents to collect.
We also analyze the most frequent web sites of the gathered document collection to see if the 
different configurations prefer different types of Web sites.

\textbf{Relevance evaluation:} In iCrawl, the topical focus of the crawl is represented 
by the crawl specification, 
a list of seeds (URLs and Social Media queries) and keywords specified by the user.
To evaluate the relevance of the crawled documents to the topical focus of the crawl, 
we build a \emph{reference vector} representing the crawl specification and \emph{document vectors} 
representing each crawled document.
Then the relevance of a crawled document is measured as its cosine similarity to the reference vector.
Such automatic evaluation is scalable and provides a fair comparison of the 
different crawler configurations.

As the reference vector is composed of multiple seed documents, 
the absolute similarity scores of any specific document to this vector is always lower than 1.
In fact, the relevance scores of the seed pages 
are in our evaluation in the interval $[0.5, 0.85]$.
%

\textbf{Freshness evaluation:} We measure the freshness as the time interval between 
fetch time of the page and the date of the page creation (see Definition~\ref{def:freshness}).
In practice, the creation date of a page is often hard to estimate because Web pages often provide 
misleading metadata. For example, the HTTP \texttt{Last-Modified} header is often equal to the 
fetch time because the page was dynamically generated.
We therefore estimate the creation date based on several content-based features. 
The features are applied sequentially until one of them finds a date.
We filter out dates before 1990 and future dates as false positives.
Table~\ref{tab:date-estimation} shows the features in the order of application 
and the percentage of documents each feature 
has been successfully applied to during the evaluation.
Using these features we could determine the date for approximately 67\% documents.
Pages for which no valid creation date could be found were excluded from the evaluation.

\begin{table}
  \centering
  \begin{tabularx}{\linewidth}{llXr}
    \toprule
    \# & ID & Feature Description & Docs \\
    \midrule
    1 & url  & date is contained in URL path &  3\% \\
    2 & time & HTML5 \texttt{<time/>} element &  9\%\\
    3 & meta & HTML \texttt{<meta/>} elements for e.g. \texttt{Last-Modified} &  8\%\\
    4 & trigger & next to trigger word such as ``updated on'' &  5\%\\
    5 & content & occurrence in text & 42\%\\
    \bottomrule
  \end{tabularx}
  \caption{Features used to estimate the page creation date in the order of application and 
  the percentage of documents.}
  \label{tab:date-estimation}
\end{table}

\subsection{Crawled Datasets} 

We perform our evaluation using two crawls on current topics: 
The \emph{\Ebola} crawl about the recent developments in the Ebola epidemy
and the \emph{\Ukraine} crawl 
about the tensions between Russia and Ukraine at the beginning of November 2014.
The crawler ran for 5 (\Ebola) resp.\@ 2 days (\Ukraine) in November 2014, 
with all configurations presented above running in parallel on separate machines.

The crawl specification for the \Ebola crawl included 
five relevant seed URLs from health organizations (\texttt{cdc.gov},
\texttt{who.int}, \texttt{healthmap.org}, \texttt{ebolacommunicationnetwork.org} and \texttt{ecdc.europa.eu}).
Twitter was queried for tweets from the users \emph{@WHO} and \emph{@Eboladeeply} and tweets containing one of the hashtags \emph{\#ebola} or \emph{\#StopEbola} or the term \emph{ebola}.
We added the keywords \emph{ebola}, \emph{liberia} and \emph{UN} for the prioritization.

For the \Ukraine crawl, we used eight seed URLs from the international
news sites including articles on the specific event (reports of tanks crossing the border 
on November 7th) as well as on the Ukraine crisis in general (\texttt{bbc.com}, \texttt{rt.com}, \texttt{kyivpost.com}, \texttt{theguardian.com}, \texttt{dw.de}, \texttt{reuters .com}, \texttt{nytimes.com} and \texttt{time.com}). 
In this crawl, Twitter was queried with the user names \emph{@KyivPost} and \emph{@Euromai\-danPR}, the hashtag \emph{\#ukraine} and the query term \emph{ukraine}.
The keywords \emph{ukraine} and \emph{russia} were added to the crawl specification for prioritization.

The crawl process was executed in batches of 1000 URLs, i.e.\@ in each batch the first 1000 URLs from the crawler queue are selected and fetched.
This batch size is a compromise between the goals of efficiency (large batch size for higher parallelism) and efficiency (small batch size for short batch processing times).
Note that this may lead to the inclusion of less relevant URLs in the beginning of the crawl when the queue contains less relevant URLs than the batch size, leading to lower precision values.

\begin{table}
  \centering
  \begin{tabular}{lrrrrrrrr}
    \toprule
    & \multicolumn{4}{c}{\Ebola} & \multicolumn{4}{c}{\Ukraine} \\
    \cmidrule(lr){2-5}
    \cmidrule(lr){6-9}
     & \TB  & \UN  & \FO  & \INT & \TB  & \UN  & \FO  & \INT \\
    \midrule
    en	     &   83 &   69 &   85 &   72 &   43 &   92 &   83 &   71 \\
    ru	     &      &      &      &      &   29 &    1 &    3 &	  17 \\
    fr	     &    2 &    6 &    3 &    4 &                           \\
    de	     &    1 &    5 &    7 &    1 &   10 &    3 &    1 &    2 \\
    zh	     &    1 &    6 &    2 &    1 &    3 &    1 &    1 &	   1 \\
    es       &    3 &    1 &    1 &    9 &                           \\
    \bottomrule
  \end{tabular}
  \caption{Language distribution (in \%) of the \Ebola and \Ukraine crawls. Values less than $1\%$ are omitted.}
  \label{tab:language-distribution}
\end{table}

The number of pages collected for the \Ebola and the \Ukraine
crawls was 16,000 and 13,800 per configuration on average, respectively.

Although the crawler started from the seed pages in English, 
it collected varying proportions of non-English content (see Table~\ref{tab:language-distribution}).
In the configurations that used Twitter, we also obtained some content from multimedia sites 
such as \texttt{instagram.com}. 
Although our manual investigation shows that the collected non-English and multimedia content is 
in many cases highly relevant to the crawl intent, automatic relevance evaluation of such content 
appears difficult for two reasons. First, the reference vector is 
built using English terms. Second, multimedia sites like \url{instagram.com} do not provide sufficient 
textual descriptions. Therefore, in this paper we exclude the content of non-English and multimedia sites
from further evaluation. We would like to investigate the issues related to the 
multilingual and multimedia collections as part of our future work.

\subsection{Web Site Distribution}
The distribution of the most frequent Web sites for each crawl is presented in Table~\ref{tab:domains}. 
For the \Ebola crawl the Social Media influenced crawlers (Twitter-based and Integrated) 
collected content most often from Social Media oriented sites like \texttt{instagram.com}, 
\texttt{linkis.com}, or \texttt{vine.co}. Also links to news aggregators 
like \texttt{newslocker.com}, \texttt{news0. tk}, \texttt{allnews24h.com}, 
and \texttt{weeder.org} were often tweeted and collected by the crawler. 
The Twitter-based crawl includes also renown addresses like \texttt{nytimes.com} 
or \texttt{huffingtonpost.com}.

The focused and unfocused crawls include most often the WHO web site (\texttt{who.int}) since it was 
part of the initial seed list. The focused crawler collected also content from CDC (\texttt{cdc.gov}) 
and a Liberian news network (\texttt{gnnliberia.com}), the Russian news agency RIA Novosti (\texttt{ria.ru}) and 
from ``Doctors Without Borders'' in Austria (\texttt{aerzte-ohne-gren\-zen.at}). The unfocused crawler 
collected instead a large number of content from Twitter and Google and the Irish news Web site ``The Journal.ie''. 

For the \Ukraine case the Twitter-based crawl behaves similar as in the \Ebola case. Not surprisingly, 
Social Media sites are most often mentioned but only one news site is among the most frequent Web sites. 
Also subdomains of the Web hoster (\texttt{hosting-test.net}) are often included as it hosts a large number 
of Ukrainian and Russian Web pages. 

All other crawlers have a high coverage of news sites like \texttt{reuters.com}, \texttt{rt.com}, \texttt{ria.ru} (both are Russian news sites), \texttt{kyivpost.com} (a Kiev newspaper) and \texttt{theguardian.com}. Furthermore, the focused and integrated crawler collected blogs posts from \texttt{wordpress.com} most frequently.

\begin{figure*}[th!]
  \centering
  \subfloat[fig:1][\Ebola Relevance]{\label{fig:batch-relevance-ebola}  \input{figures/batch_relevance}}
  \subfloat[fig:2][\Ukraine Relevance]{\label{fig:batch-relevance-ukraine}\input{figures/batch_relevance_ukraine2}}\\
   \subfloat[fig:3][\Ebola Freshness]{\label{fig:batch-freshness-ebola}\input{figures/batch_freshness}}
  \subfloat[fig:4][\Ukraine Freshness]{\label{fig:batch-freshness-ukraine}\input{figures/batch_freshness_ukraine2}}\\
   \caption{Relevance and freshness of the documents during the \Ebola and 
   \Ukraine crawls. The $X$-axis represents the 
  batches processed by the crawler in the chronological order. The $Y$-axis represents 
  the average relevance and freshness (in hours) of the documents in a batch, respectively.
  Each curve corresponds to a crawler configuration (Focused (FO), Twitter-based (TB), 
  and Integrated (INT)). The unfocused crawler performed worse than the other baselines and was therefore omitted for readability. Higher relevance values and lower freshness values are better.}
  \label{fig:batch-relevance}
\end{figure*}
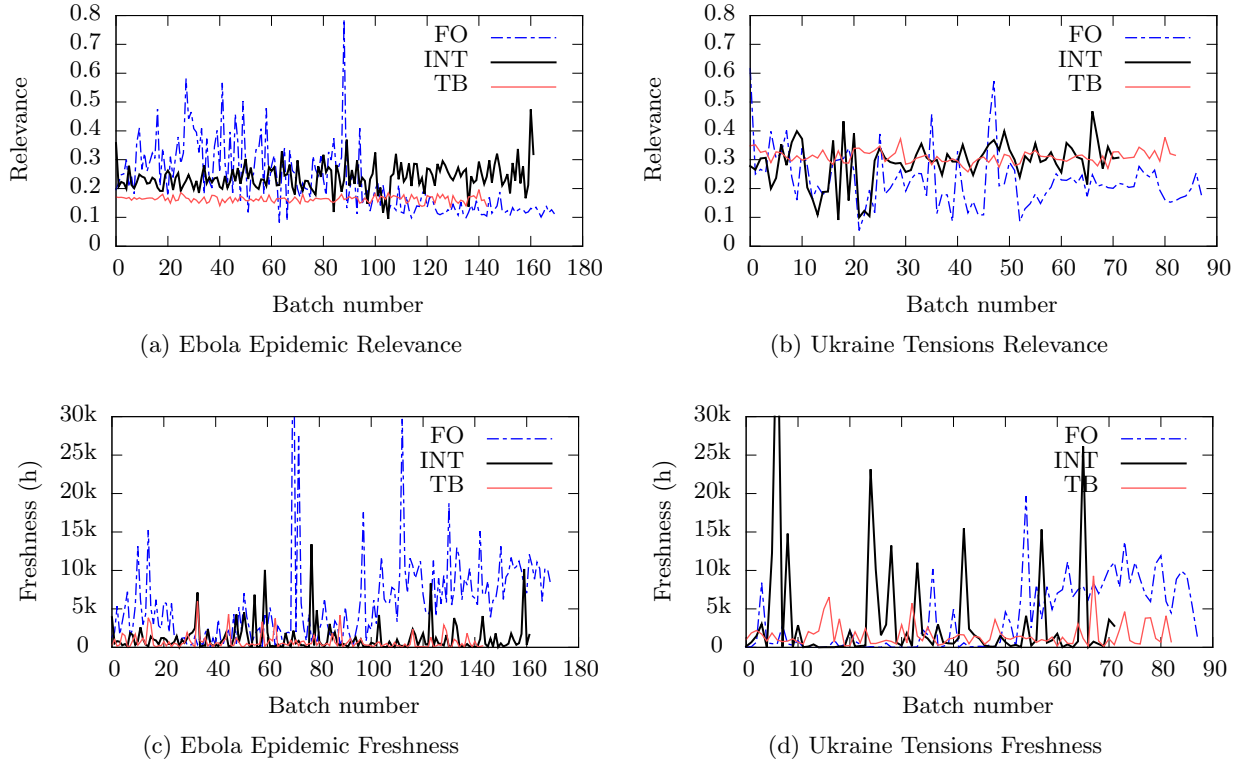

\begin{table*}
\centering{}
\input{input/tab-domains}
\caption{Most frequent Web sites in each crawl configuration for the \Ebola and the \Ukraine crawl.}
\label{tab:domains}
\end{table*}

\subsection{Relevance Evaluation}

Figure~\ref{fig:batch-relevance-ebola} and Figure~\ref{fig:batch-relevance-ukraine} present the 
relevance of the documents collected during the \Ebola crawl and the \Ukraine crawl, respectively. 
The $X$-axis reflects the batches downloaded by the crawler in chronological order.
The $Y$-axis represents the average relevance of the 
documents in a batch (higher values correspond to more relevant documents).

\textbf{Relevance of the \Ebola Crawl:}
For the \Ebola crawl shown in Figure~\ref{fig:batch-relevance-ebola} we can observe 
at the beginning of the crawling process that the average relevance of the focused crawler is higher than for the other configurations. This can be attributed to the high relevance of the 
seed pages and the high effectiveness of the focusing strategy. 
In some batches the focused crawler downloaded a large number of highly relevant pages at once, \eg from the WHO website, which are visible as spikes in the average relevance (\eg in batch 92).

However, Figure~\ref{fig:batch-relevance-ebola} also indicates that the average 
relevance of the content collected by the focused crawler drops over time.
This can be explained by the limited number of relevant web resources connected to the initial seed URLs.

In contrast, the average relevance of the Twitter-based crawl remains at a lower but more stable level over time. 
The reason for the on average lower relevance scores of the Twitter-based crawl can be explained by the source of the collected documents:
This crawler collects more `popular' documents, \eg from Social Media sites (see Table~\ref{tab:domains}), which use a different vocabulary than the seed documents that were used to create the reference vector.

%

The Twitter input enables the integrated crawler to find relevant content independently 
of the original seed URLs and thus to remain more stable over time with respect to the relevance 
compared to the focused crawler. 
The focusing of the integrated crawler can handle the noisy input from Twitter which can be observed when it starts to outperform 
the baseline focused crawler after around 100 batches in the \Ebola crawl.

\textbf{Relevance of the \Ukraine Crawl:}
In case of the \Ukraine crawl the general observations remain similar as shown in 
Figure~\ref{fig:batch-relevance-ukraine}. The focused crawler begins with highly relevant content 
due to the impact of the seed list. Afterwards it drops rapidly and continues with quite 
high variance until batch 50. Finally, the variance decreases and the content relevance remains 
on a lower level.

In contrast, the Twitter-based \Ukraine crawl shows similar to the \Ebola case a stable relevance over 
the entire crawl duration. Due to the closeness of the crawl to the ongoing event, 
the crawled content is of higher relevance compared to the \Ebola case. 

The integrated crawler shows in the early stages a similar high variance as the focused crawler. 
The reason is again the influence of the original seeds at the beginning of the crawl. 
Later it drops but stabilizes after 20 batches due to the increasing impact of the extracted links from Twitter. 
Over time the integrated crawler demonstrates a similar relevance as the Twitter-based crawler 
but with a higher variance. 

\textbf{Relevance Summary:}
In summary, in  both use cases we can observe the positive influence of the Twitter integration 
on the relevance distribution over time. Although the baseline focused crawler can 
obtain relevant pages at the beginning, its ability to identify such pages highly depends on the seeds
and reduces over time. 
In contrast, continuous Twitter input enables the integrated crawler 
to outperform the baseline focused crawler as the crawl progresses.

The high variance of the focused and integrated crawlers can be explained by the number of links \eg from site menus even of relevant pages.
Those still need to be fetched to realize that they are irrelevant.
The number of those links varies and therefore causes a wider spread of the average relevance per crawler batch.
In case of the Twitter-based crawl, the variance is significantly lower. 
This is because in this crawl we just follow the already filtered links from the Tweets, but do not
collect the pages from the Web graph around them. However, the cost of the lower variance 
is the lower coverage of the related pages on the Web. 

We also performed an evaluation of relevance and freshness with the unfocused crawler baseline. As expected, 
this configuration was outperformed by the focused crawler baseline with respect to both relevance and freshness.
We omit this configuration from the graphs in Figure~\ref{fig:batch-relevance} for readability.

\subsection{Document Freshness Evaluation}

Figure~\ref{fig:batch-freshness-ebola} and Figure~\ref{fig:batch-freshness-ukraine} show the freshness of the documents collected during the \Ebola and \Ukraine crawls, respectively. The X-axis reflects the batches downloaded by the crawler over time.
The Y-axis show the average freshness of the documents in a batch in hours (lower freshness values 
correspond to the more recent documents).

\textbf{Freshness of the \Ebola Crawl:}
As we can observe in Figure~\ref{fig:batch-freshness-ebola}, the average freshness of the Twitter-based 
crawler is the best throughout the crawling process, followed by the integrated crawler. 
The focused crawler fetches more older pages after about 100 batches. 
This is similar to the trend we observed for relevance, where the focused crawler collected less relevant content after a certain number of pages.

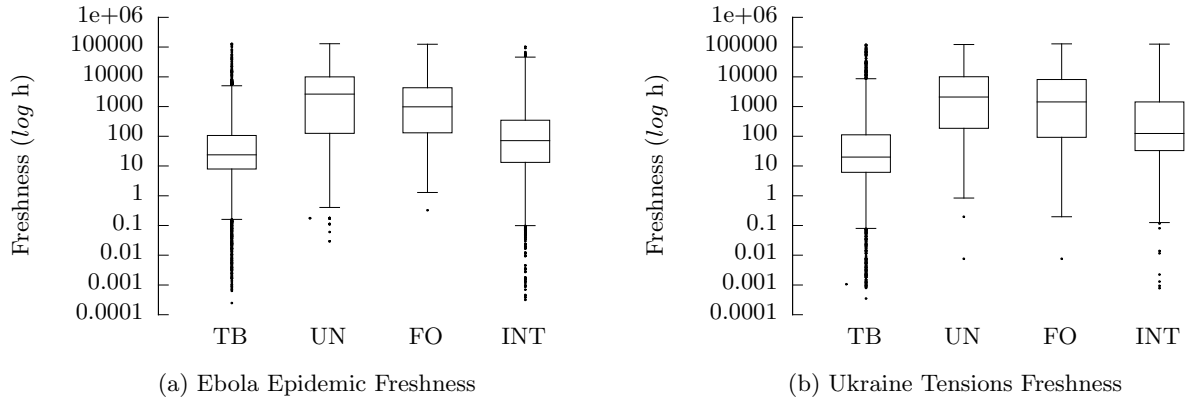
\begin{figure*}[th!]
\vspace{-.5em}
  \centering
  \subfloat[fig:1][\Ebola Freshness]{\label{fig:freshness-distribution-ebola}  \input{figures/freshness_distribution}}
  \subfloat[fig:2][\Ukraine Freshness]{\label{fig:freshness-distribution-ukraine}\input{figures/freshness_distribution_ukraine2}}\\
   \caption{The boxplot representing the distribution of the freshness in the \Ebola and 
   \Ukraine crawls obtained with different crawler configurations. 
   The $Y$-axis (log scale) presents the freshness of the content (in hours).
   The box boundaries correspond to the upper and lower quartile of the data points.}
  \label{fig:freshness-distribution}
\end{figure*}

Figure \ref{fig:freshness-distribution} shows the distribution of the freshness obtained by 
different crawler configurations on the topic of \Ebola. The $Y$-axis  represents 
the freshness of the crawled content (in hours). The box boundaries correspond to the upper and lower 
quartile of the data points, such that 50\% of the data points lay inside the boxes. 

The Twitter-driven settings achieve a significantly better freshness than the other crawl configurations, as shown in Figure \ref{fig:freshness-distribution-ebola}
For example, 
the Twitter-based crawl has the best freshness values with a median of 24~hours, second 
is the integrated crawler  with a median freshness of 65~hours. 
The focused crawler collects pages with the median freshness of 980~hours, which is 15 times longer 
than the integrated crawl, while the median of the unfocused crawler is even 2300~hours (i.e. 
approximately 3 months). 

\textbf{Freshness of the \Ukraine Crawl:}
Figure~\ref{fig:batch-freshness-ukraine} shows the freshness of the collected pages in the \Ukraine crawl. 
The Twitter-based crawl shows again the highest freshness. 

For the focused crawl we see the same behavior as for the \Ebola crawl: The freshness is high at the beginning, but starts decreasing after 50 batches.
Again this is accompanied by a drop in relevance: the crawler seems to have collected all reachable relevant content here as well.

The content crawled by the integrated crawler is in general of similar and sometimes higher 
freshness as that of the Twitter-based crawler.
However, it has some outlier batches where more old content is crawled.
As discussed in the relevance of the \Ukraine crawl this can be another indicator that old and 
unrelated content has been collected before the crawler was able to follow related and fresh links.

These observations are confirmed by the distribution of the freshness shown in Figure~\ref{fig:freshness-distribution-ukraine}. 
Again the values differ significantly across the crawler configurations. 
The Twitter-based crawl provides content with the highest freshness at a median of 20~hours. The 
freshness values of the unfocused and focused crawler are rather low with a 
median of 1920 and 1415~hours, respectively. Finally, the integrated crawler results in a mixture of 
fresh and old content with the median of 120~hours and therefore with 
higher degree of fresh content compared to both of the unfocused and focused baselines.

\textbf{Freshness Summary:}
Overall, we can conclude that the integrated crawler significantly outperforms both unfocused 
and focused baseline crawlers in terms of freshness, especially as the crawler moves away from the (fresh) seed URLs. The Twitter-based crawls demonstrate the highest freshness, 
which shows that the Twitter input clearly contributes to the improved freshness of the 
integrated crawler.

\subsection{Evaluation Results Discussion}

As our evaluation results indicate, the most effective crawler configuration 
depends on the distribution of the relevant data across the media. For example,
the \Ebola topic is more generic and is covered widely in news sites. 
However, Tweets containing keywords related to Ebola address a variety 
of aspects and are often off-topic for the crawl intent (e.g. ``Things worse than ebola'').
In contrast, the \Ukraine topic targeted a very recent 
and specific event. For this topic we observe higher relevance 
of the Tweets and lower coverage in the news sites.

Our experiments show that the \Ebola topic profits most from focused Web crawling, but the baseline focused crawler still exhaust its seed URLs very quickly. On the other hand, 
the \Ukraine topic profits most from the relevant Tweets, here Twitter alone however
 provides only the limited view of what Twitter users decide to share.
 The integrated crawler takes advantage of both precise focused crawling and 
 continuous input of fresh Twitter stream and automatically adapts
 its behavior towards the most promising information source.
 



%% file: figures/batch_relevance.tex
\begingroup
  \makeatletter
  \providecommand\color[2][]{%
    \GenericError{(gnuplot) \space\space\space\@spaces}{%
      Package color not loaded in conjunction with
      terminal option `colourtext'%
    }{See the gnuplot documentation for explanation.%
    }{Either use 'blacktext' in gnuplot or load the package
      color.sty in LaTeX.}%
    \renewcommand\color[2][]{}%
  }%
  \providecommand\includegraphics[2][]{%
    \GenericError{(gnuplot) \space\space\space\@spaces}{%
      Package graphicx or graphics not loaded%
    }{See the gnuplot documentation for explanation.%
    }{The gnuplot epslatex terminal needs graphicx.sty or graphics.sty.}%
    \renewcommand\includegraphics[2][]{}%
  }%
  \providecommand\rotatebox[2]{#2}%
  \@ifundefined{ifGPcolor}{%
    \newif\ifGPcolor
    \GPcolortrue
  }{}%
  \@ifundefined{ifGPblacktext}{%
    \newif\ifGPblacktext
    \GPblacktexttrue
  }{}%
  \let\gplgaddtomacro\g@addto@macro
  \gdef\gplbacktext{}%
  \gdef\gplfronttext{}%
  \makeatother
  \ifGPblacktext
    \def\colorrgb#1{}%
    \def\colorgray#1{}%
  \else
    \ifGPcolor
      \def\colorrgb#1{\color[rgb]{#1}}%
      \def\colorgray#1{\color[gray]{#1}}%
      \expandafter\def\csname LTw\endcsname{\color{white}}%
      \expandafter\def\csname LTb\endcsname{\color{black}}%
      \expandafter\def\csname LTa\endcsname{\color{black}}%
      \expandafter\def\csname LT0\endcsname{\color[rgb]{1,0,0}}%
      \expandafter\def\csname LT1\endcsname{\color[rgb]{0,1,0}}%
      \expandafter\def\csname LT2\endcsname{\color[rgb]{0,0,1}}%
      \expandafter\def\csname LT3\endcsname{\color[rgb]{1,0,1}}%
      \expandafter\def\csname LT4\endcsname{\color[rgb]{0,1,1}}%
      \expandafter\def\csname LT5\endcsname{\color[rgb]{1,1,0}}%
      \expandafter\def\csname LT6\endcsname{\color[rgb]{0,0,0}}%
      \expandafter\def\csname LT7\endcsname{\color[rgb]{1,0.3,0}}%
      \expandafter\def\csname LT8\endcsname{\color[rgb]{0.5,0.5,0.5}}%
    \else
      \def\colorrgb#1{\color{black}}%
      \def\colorgray#1{\color[gray]{#1}}%
      \expandafter\def\csname LTw\endcsname{\color{white}}%
      \expandafter\def\csname LTb\endcsname{\color{black}}%
      \expandafter\def\csname LTa\endcsname{\color{black}}%
      \expandafter\def\csname LT0\endcsname{\color{black}}%
      \expandafter\def\csname LT1\endcsname{\color{black}}%
      \expandafter\def\csname LT2\endcsname{\color{black}}%
      \expandafter\def\csname LT3\endcsname{\color{black}}%
      \expandafter\def\csname LT4\endcsname{\color{black}}%
      \expandafter\def\csname LT5\endcsname{\color{black}}%
      \expandafter\def\csname LT6\endcsname{\color{black}}%
      \expandafter\def\csname LT7\endcsname{\color{black}}%
      \expandafter\def\csname LT8\endcsname{\color{black}}%
    \fi
  \fi
  \setlength{\unitlength}{0.0500bp}%
  \begin{picture}(4780.00,2540.00)%
    \gplgaddtomacro\gplbacktext{%
      \csname LTb\endcsname%
      \put(854,595){\makebox(0,0)[r]{\strut{} 0}}%
      \csname LTb\endcsname%
      \put(854,813){\makebox(0,0)[r]{\strut{} 0.1}}%
      \csname LTb\endcsname%
      \put(854,1030){\makebox(0,0)[r]{\strut{} 0.2}}%
      \csname LTb\endcsname%
      \put(854,1248){\makebox(0,0)[r]{\strut{} 0.3}}%
      \csname LTb\endcsname%
      \put(854,1465){\makebox(0,0)[r]{\strut{} 0.4}}%
      \csname LTb\endcsname%
      \put(854,1683){\makebox(0,0)[r]{\strut{} 0.5}}%
      \csname LTb\endcsname%
      \put(854,1900){\makebox(0,0)[r]{\strut{} 0.6}}%
      \csname LTb\endcsname%
      \put(854,2118){\makebox(0,0)[r]{\strut{} 0.7}}%
      \csname LTb\endcsname%
      \put(854,2335){\makebox(0,0)[r]{\strut{} 0.8}}%
      \csname LTb\endcsname%
      \put(956,409){\makebox(0,0){\strut{} 0}}%
      \csname LTb\endcsname%
      \put(1347,409){\makebox(0,0){\strut{} 20}}%
      \csname LTb\endcsname%
      \put(1738,409){\makebox(0,0){\strut{} 40}}%
      \csname LTb\endcsname%
      \put(2128,409){\makebox(0,0){\strut{} 60}}%
      \csname LTb\endcsname%
      \put(2519,409){\makebox(0,0){\strut{} 80}}%
      \csname LTb\endcsname%
      \put(2910,409){\makebox(0,0){\strut{} 100}}%
      \csname LTb\endcsname%
      \put(3301,409){\makebox(0,0){\strut{} 120}}%
      \csname LTb\endcsname%
      \put(3691,409){\makebox(0,0){\strut{} 140}}%
      \csname LTb\endcsname%
      \put(4082,409){\makebox(0,0){\strut{} 160}}%
      \csname LTb\endcsname%
      \put(4473,409){\makebox(0,0){\strut{} 180}}%
      \csname LTb\endcsname%
      \put(251,1465){\rotatebox{-270}{\makebox(0,0){\strut{}Relevance}}}%
      \csname LTb\endcsname%
      \put(2714,130){\makebox(0,0){\strut{}Batch number}}%
    }%
    \gplgaddtomacro\gplfronttext{%
      \csname LTb\endcsname%
      \put(3685,2168){\makebox(0,0)[r]{\strut{}\FO}}%
      \csname LTb\endcsname%
      \put(3685,1982){\makebox(0,0)[r]{\strut{}\INT}}%
      \csname LTb\endcsname%
      \put(3685,1796){\makebox(0,0)[r]{\strut{}\TB}}%
    }%
    \gplbacktext
    \put(0,0){\includegraphics{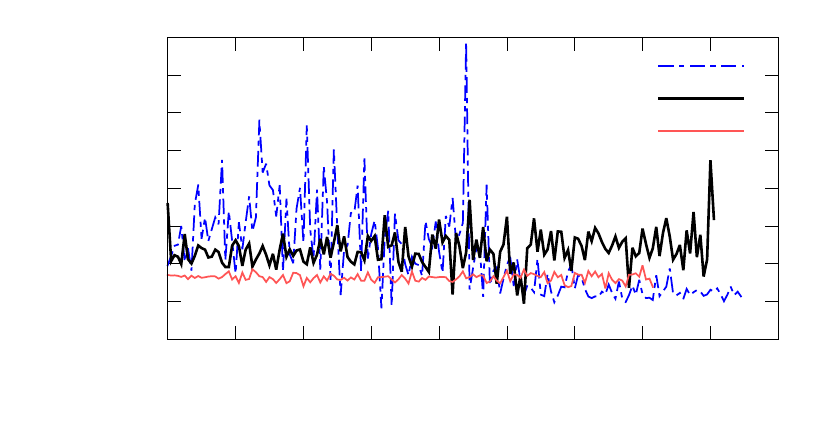}}%
    \gplfronttext
  \end{picture}%
\endgroup

%% file: figures/batch_relevance_ukraine2.tex
\begingroup
  \makeatletter
  \providecommand\color[2][]{%
    \GenericError{(gnuplot) \space\space\space\@spaces}{%
      Package color not loaded in conjunction with
      terminal option `colourtext'%
    }{See the gnuplot documentation for explanation.%
    }{Either use 'blacktext' in gnuplot or load the package
      color.sty in LaTeX.}%
    \renewcommand\color[2][]{}%
  }%
  \providecommand\includegraphics[2][]{%
    \GenericError{(gnuplot) \space\space\space\@spaces}{%
      Package graphicx or graphics not loaded%
    }{See the gnuplot documentation for explanation.%
    }{The gnuplot epslatex terminal needs graphicx.sty or graphics.sty.}%
    \renewcommand\includegraphics[2][]{}%
  }%
  \providecommand\rotatebox[2]{#2}%
  \@ifundefined{ifGPcolor}{%
    \newif\ifGPcolor
    \GPcolortrue
  }{}%
  \@ifundefined{ifGPblacktext}{%
    \newif\ifGPblacktext
    \GPblacktexttrue
  }{}%
  \let\gplgaddtomacro\g@addto@macro
  \gdef\gplbacktext{}%
  \gdef\gplfronttext{}%
  \makeatother
  \ifGPblacktext
    \def\colorrgb#1{}%
    \def\colorgray#1{}%
  \else
    \ifGPcolor
      \def\colorrgb#1{\color[rgb]{#1}}%
      \def\colorgray#1{\color[gray]{#1}}%
      \expandafter\def\csname LTw\endcsname{\color{white}}%
      \expandafter\def\csname LTb\endcsname{\color{black}}%
      \expandafter\def\csname LTa\endcsname{\color{black}}%
      \expandafter\def\csname LT0\endcsname{\color[rgb]{1,0,0}}%
      \expandafter\def\csname LT1\endcsname{\color[rgb]{0,1,0}}%
      \expandafter\def\csname LT2\endcsname{\color[rgb]{0,0,1}}%
      \expandafter\def\csname LT3\endcsname{\color[rgb]{1,0,1}}%
      \expandafter\def\csname LT4\endcsname{\color[rgb]{0,1,1}}%
      \expandafter\def\csname LT5\endcsname{\color[rgb]{1,1,0}}%
      \expandafter\def\csname LT6\endcsname{\color[rgb]{0,0,0}}%
      \expandafter\def\csname LT7\endcsname{\color[rgb]{1,0.3,0}}%
      \expandafter\def\csname LT8\endcsname{\color[rgb]{0.5,0.5,0.5}}%
    \else
      \def\colorrgb#1{\color{black}}%
      \def\colorgray#1{\color[gray]{#1}}%
      \expandafter\def\csname LTw\endcsname{\color{white}}%
      \expandafter\def\csname LTb\endcsname{\color{black}}%
      \expandafter\def\csname LTa\endcsname{\color{black}}%
      \expandafter\def\csname LT0\endcsname{\color{black}}%
      \expandafter\def\csname LT1\endcsname{\color{black}}%
      \expandafter\def\csname LT2\endcsname{\color{black}}%
      \expandafter\def\csname LT3\endcsname{\color{black}}%
      \expandafter\def\csname LT4\endcsname{\color{black}}%
      \expandafter\def\csname LT5\endcsname{\color{black}}%
      \expandafter\def\csname LT6\endcsname{\color{black}}%
      \expandafter\def\csname LT7\endcsname{\color{black}}%
      \expandafter\def\csname LT8\endcsname{\color{black}}%
    \fi
  \fi
  \setlength{\unitlength}{0.0500bp}%
  \begin{picture}(4780.00,2540.00)%
    \gplgaddtomacro\gplbacktext{%
      \csname LTb\endcsname%
      \put(854,595){\makebox(0,0)[r]{\strut{} 0}}%
      \csname LTb\endcsname%
      \put(854,813){\makebox(0,0)[r]{\strut{} 0.1}}%
      \csname LTb\endcsname%
      \put(854,1030){\makebox(0,0)[r]{\strut{} 0.2}}%
      \csname LTb\endcsname%
      \put(854,1248){\makebox(0,0)[r]{\strut{} 0.3}}%
      \csname LTb\endcsname%
      \put(854,1465){\makebox(0,0)[r]{\strut{} 0.4}}%
      \csname LTb\endcsname%
      \put(854,1683){\makebox(0,0)[r]{\strut{} 0.5}}%
      \csname LTb\endcsname%
      \put(854,1900){\makebox(0,0)[r]{\strut{} 0.6}}%
      \csname LTb\endcsname%
      \put(854,2118){\makebox(0,0)[r]{\strut{} 0.7}}%
      \csname LTb\endcsname%
      \put(854,2335){\makebox(0,0)[r]{\strut{} 0.8}}%
      \csname LTb\endcsname%
      \put(956,409){\makebox(0,0){\strut{} 0}}%
      \csname LTb\endcsname%
      \put(1347,409){\makebox(0,0){\strut{} 10}}%
      \csname LTb\endcsname%
      \put(1738,409){\makebox(0,0){\strut{} 20}}%
      \csname LTb\endcsname%
      \put(2128,409){\makebox(0,0){\strut{} 30}}%
      \csname LTb\endcsname%
      \put(2519,409){\makebox(0,0){\strut{} 40}}%
      \csname LTb\endcsname%
      \put(2910,409){\makebox(0,0){\strut{} 50}}%
      \csname LTb\endcsname%
      \put(3301,409){\makebox(0,0){\strut{} 60}}%
      \csname LTb\endcsname%
      \put(3691,409){\makebox(0,0){\strut{} 70}}%
      \csname LTb\endcsname%
      \put(4082,409){\makebox(0,0){\strut{} 80}}%
      \csname LTb\endcsname%
      \put(4473,409){\makebox(0,0){\strut{} 90}}%
      \csname LTb\endcsname%
      \put(251,1465){\rotatebox{-270}{\makebox(0,0){\strut{}Relevance}}}%
      \csname LTb\endcsname%
      \put(2714,130){\makebox(0,0){\strut{}Batch number}}%
    }%
    \gplgaddtomacro\gplfronttext{%
      \csname LTb\endcsname%
      \put(3685,2168){\makebox(0,0)[r]{\strut{}\FO}}%
      \csname LTb\endcsname%
      \put(3685,1982){\makebox(0,0)[r]{\strut{}\INT}}%
      \csname LTb\endcsname%
      \put(3685,1796){\makebox(0,0)[r]{\strut{}\TB}}%
    }%
    \gplbacktext
    \put(0,0){\includegraphics{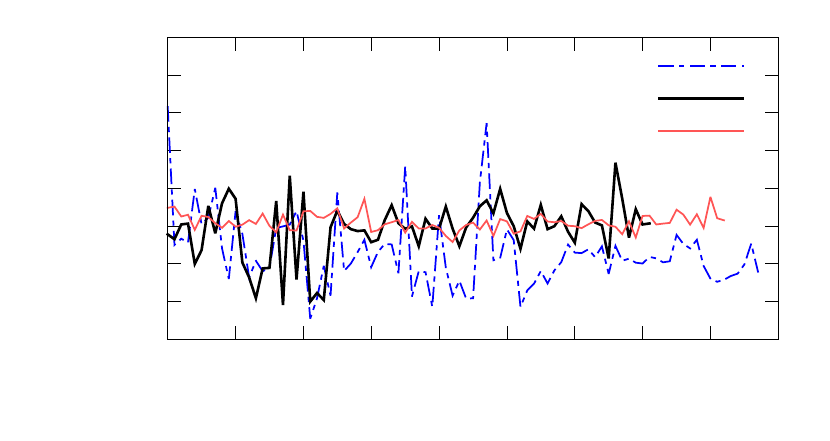}}%
    \gplfronttext
  \end{picture}%
\endgroup

%% file: figures/batch_freshness.tex
\begingroup
  \makeatletter
  \providecommand\color[2][]{%
    \GenericError{(gnuplot) \space\space\space\@spaces}{%
      Package color not loaded in conjunction with
      terminal option `colourtext'%
    }{See the gnuplot documentation for explanation.%
    }{Either use 'blacktext' in gnuplot or load the package
      color.sty in LaTeX.}%
    \renewcommand\color[2][]{}%
  }%
  \providecommand\includegraphics[2][]{%
    \GenericError{(gnuplot) \space\space\space\@spaces}{%
      Package graphicx or graphics not loaded%
    }{See the gnuplot documentation for explanation.%
    }{The gnuplot epslatex terminal needs graphicx.sty or graphics.sty.}%
    \renewcommand\includegraphics[2][]{}%
  }%
  \providecommand\rotatebox[2]{#2}%
  \@ifundefined{ifGPcolor}{%
    \newif\ifGPcolor
    \GPcolortrue
  }{}%
  \@ifundefined{ifGPblacktext}{%
    \newif\ifGPblacktext
    \GPblacktexttrue
  }{}%
  \let\gplgaddtomacro\g@addto@macro
  \gdef\gplbacktext{}%
  \gdef\gplfronttext{}%
  \makeatother
  \ifGPblacktext
    \def\colorrgb#1{}%
    \def\colorgray#1{}%
  \else
    \ifGPcolor
      \def\colorrgb#1{\color[rgb]{#1}}%
      \def\colorgray#1{\color[gray]{#1}}%
      \expandafter\def\csname LTw\endcsname{\color{white}}%
      \expandafter\def\csname LTb\endcsname{\color{black}}%
      \expandafter\def\csname LTa\endcsname{\color{black}}%
      \expandafter\def\csname LT0\endcsname{\color[rgb]{1,0,0}}%
      \expandafter\def\csname LT1\endcsname{\color[rgb]{0,1,0}}%
      \expandafter\def\csname LT2\endcsname{\color[rgb]{0,0,1}}%
      \expandafter\def\csname LT3\endcsname{\color[rgb]{1,0,1}}%
      \expandafter\def\csname LT4\endcsname{\color[rgb]{0,1,1}}%
      \expandafter\def\csname LT5\endcsname{\color[rgb]{1,1,0}}%
      \expandafter\def\csname LT6\endcsname{\color[rgb]{0,0,0}}%
      \expandafter\def\csname LT7\endcsname{\color[rgb]{1,0.3,0}}%
      \expandafter\def\csname LT8\endcsname{\color[rgb]{0.5,0.5,0.5}}%
    \else
      \def\colorrgb#1{\color{black}}%
      \def\colorgray#1{\color[gray]{#1}}%
      \expandafter\def\csname LTw\endcsname{\color{white}}%
      \expandafter\def\csname LTb\endcsname{\color{black}}%
      \expandafter\def\csname LTa\endcsname{\color{black}}%
      \expandafter\def\csname LT0\endcsname{\color{black}}%
      \expandafter\def\csname LT1\endcsname{\color{black}}%
      \expandafter\def\csname LT2\endcsname{\color{black}}%
      \expandafter\def\csname LT3\endcsname{\color{black}}%
      \expandafter\def\csname LT4\endcsname{\color{black}}%
      \expandafter\def\csname LT5\endcsname{\color{black}}%
      \expandafter\def\csname LT6\endcsname{\color{black}}%
      \expandafter\def\csname LT7\endcsname{\color{black}}%
      \expandafter\def\csname LT8\endcsname{\color{black}}%
    \fi
  \fi
  \setlength{\unitlength}{0.0500bp}%
  \begin{picture}(4780.00,2540.00)%
    \gplgaddtomacro\gplbacktext{%
      \csname LTb\endcsname%
      \put(854,595){\makebox(0,0)[r]{\strut{}0 }}%
      \csname LTb\endcsname%
      \put(854,885){\makebox(0,0)[r]{\strut{}5k}}%
      \csname LTb\endcsname%
      \put(854,1175){\makebox(0,0)[r]{\strut{}10k}}%
      \csname LTb\endcsname%
      \put(854,1465){\makebox(0,0)[r]{\strut{}15k}}%
      \csname LTb\endcsname%
      \put(854,1755){\makebox(0,0)[r]{\strut{}20k}}%
      \csname LTb\endcsname%
      \put(854,2045){\makebox(0,0)[r]{\strut{}25k}}%
      \csname LTb\endcsname%
      \put(854,2335){\makebox(0,0)[r]{\strut{}30k}}%
      \csname LTb\endcsname%
      \put(956,409){\makebox(0,0){\strut{} 0}}%
      \csname LTb\endcsname%
      \put(1347,409){\makebox(0,0){\strut{} 20}}%
      \csname LTb\endcsname%
      \put(1738,409){\makebox(0,0){\strut{} 40}}%
      \csname LTb\endcsname%
      \put(2128,409){\makebox(0,0){\strut{} 60}}%
      \csname LTb\endcsname%
      \put(2519,409){\makebox(0,0){\strut{} 80}}%
      \csname LTb\endcsname%
      \put(2910,409){\makebox(0,0){\strut{} 100}}%
      \csname LTb\endcsname%
      \put(3301,409){\makebox(0,0){\strut{} 120}}%
      \csname LTb\endcsname%
      \put(3691,409){\makebox(0,0){\strut{} 140}}%
      \csname LTb\endcsname%
      \put(4082,409){\makebox(0,0){\strut{} 160}}%
      \csname LTb\endcsname%
      \put(4473,409){\makebox(0,0){\strut{} 180}}%
      \csname LTb\endcsname%
      \put(353,1465){\rotatebox{-270}{\makebox(0,0){\strut{}Freshness (h)}}}%
      \csname LTb\endcsname%
      \put(2714,130){\makebox(0,0){\strut{}Batch number}}%
    }%
    \gplgaddtomacro\gplfronttext{%
      \csname LTb\endcsname%
      \put(3685,2168){\makebox(0,0)[r]{\strut{}\FO}}%
      \csname LTb\endcsname%
      \put(3685,1982){\makebox(0,0)[r]{\strut{}\INT}}%
      \csname LTb\endcsname%
      \put(3685,1796){\makebox(0,0)[r]{\strut{}\TB}}%
    }%
    \gplbacktext
    \put(0,0){\includegraphics{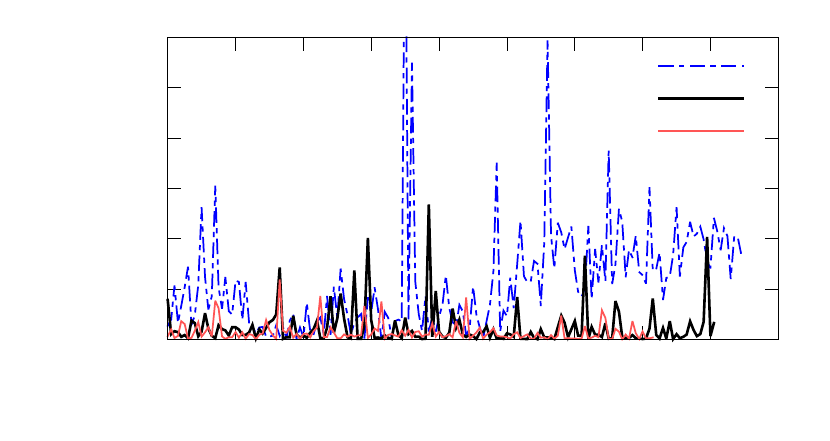}}%
    \gplfronttext
  \end{picture}%
\endgroup

%% file: figures/batch_freshness_ukraine2.tex
\begingroup
  \makeatletter
  \providecommand\color[2][]{%
    \GenericError{(gnuplot) \space\space\space\@spaces}{%
      Package color not loaded in conjunction with
      terminal option `colourtext'%
    }{See the gnuplot documentation for explanation.%
    }{Either use 'blacktext' in gnuplot or load the package
      color.sty in LaTeX.}%
    \renewcommand\color[2][]{}%
  }%
  \providecommand\includegraphics[2][]{%
    \GenericError{(gnuplot) \space\space\space\@spaces}{%
      Package graphicx or graphics not loaded%
    }{See the gnuplot documentation for explanation.%
    }{The gnuplot epslatex terminal needs graphicx.sty or graphics.sty.}%
    \renewcommand\includegraphics[2][]{}%
  }%
  \providecommand\rotatebox[2]{#2}%
  \@ifundefined{ifGPcolor}{%
    \newif\ifGPcolor
    \GPcolortrue
  }{}%
  \@ifundefined{ifGPblacktext}{%
    \newif\ifGPblacktext
    \GPblacktexttrue
  }{}%
  \let\gplgaddtomacro\g@addto@macro
  \gdef\gplbacktext{}%
  \gdef\gplfronttext{}%
  \makeatother
  \ifGPblacktext
    \def\colorrgb#1{}%
    \def\colorgray#1{}%
  \else
    \ifGPcolor
      \def\colorrgb#1{\color[rgb]{#1}}%
      \def\colorgray#1{\color[gray]{#1}}%
      \expandafter\def\csname LTw\endcsname{\color{white}}%
      \expandafter\def\csname LTb\endcsname{\color{black}}%
      \expandafter\def\csname LTa\endcsname{\color{black}}%
      \expandafter\def\csname LT0\endcsname{\color[rgb]{1,0,0}}%
      \expandafter\def\csname LT1\endcsname{\color[rgb]{0,1,0}}%
      \expandafter\def\csname LT2\endcsname{\color[rgb]{0,0,1}}%
      \expandafter\def\csname LT3\endcsname{\color[rgb]{1,0,1}}%
      \expandafter\def\csname LT4\endcsname{\color[rgb]{0,1,1}}%
      \expandafter\def\csname LT5\endcsname{\color[rgb]{1,1,0}}%
      \expandafter\def\csname LT6\endcsname{\color[rgb]{0,0,0}}%
      \expandafter\def\csname LT7\endcsname{\color[rgb]{1,0.3,0}}%
      \expandafter\def\csname LT8\endcsname{\color[rgb]{0.5,0.5,0.5}}%
    \else
      \def\colorrgb#1{\color{black}}%
      \def\colorgray#1{\color[gray]{#1}}%
      \expandafter\def\csname LTw\endcsname{\color{white}}%
      \expandafter\def\csname LTb\endcsname{\color{black}}%
      \expandafter\def\csname LTa\endcsname{\color{black}}%
      \expandafter\def\csname LT0\endcsname{\color{black}}%
      \expandafter\def\csname LT1\endcsname{\color{black}}%
      \expandafter\def\csname LT2\endcsname{\color{black}}%
      \expandafter\def\csname LT3\endcsname{\color{black}}%
      \expandafter\def\csname LT4\endcsname{\color{black}}%
      \expandafter\def\csname LT5\endcsname{\color{black}}%
      \expandafter\def\csname LT6\endcsname{\color{black}}%
      \expandafter\def\csname LT7\endcsname{\color{black}}%
      \expandafter\def\csname LT8\endcsname{\color{black}}%
    \fi
  \fi
  \setlength{\unitlength}{0.0500bp}%
  \begin{picture}(4780.00,2540.00)%
    \gplgaddtomacro\gplbacktext{%
      \csname LTb\endcsname%
      \put(854,595){\makebox(0,0)[r]{\strut{}0 }}%
      \csname LTb\endcsname%
      \put(854,885){\makebox(0,0)[r]{\strut{}5k}}%
      \csname LTb\endcsname%
      \put(854,1175){\makebox(0,0)[r]{\strut{}10k}}%
      \csname LTb\endcsname%
      \put(854,1465){\makebox(0,0)[r]{\strut{}15k}}%
      \csname LTb\endcsname%
      \put(854,1755){\makebox(0,0)[r]{\strut{}20k}}%
      \csname LTb\endcsname%
      \put(854,2045){\makebox(0,0)[r]{\strut{}25k}}%
      \csname LTb\endcsname%
      \put(854,2335){\makebox(0,0)[r]{\strut{}30k}}%
      \csname LTb\endcsname%
      \put(956,409){\makebox(0,0){\strut{} 0}}%
      \csname LTb\endcsname%
      \put(1347,409){\makebox(0,0){\strut{} 10}}%
      \csname LTb\endcsname%
      \put(1738,409){\makebox(0,0){\strut{} 20}}%
      \csname LTb\endcsname%
      \put(2128,409){\makebox(0,0){\strut{} 30}}%
      \csname LTb\endcsname%
      \put(2519,409){\makebox(0,0){\strut{} 40}}%
      \csname LTb\endcsname%
      \put(2910,409){\makebox(0,0){\strut{} 50}}%
      \csname LTb\endcsname%
      \put(3301,409){\makebox(0,0){\strut{} 60}}%
      \csname LTb\endcsname%
      \put(3691,409){\makebox(0,0){\strut{} 70}}%
      \csname LTb\endcsname%
      \put(4082,409){\makebox(0,0){\strut{} 80}}%
      \csname LTb\endcsname%
      \put(4473,409){\makebox(0,0){\strut{} 90}}%
      \csname LTb\endcsname%
      \put(353,1465){\rotatebox{-270}{\makebox(0,0){\strut{}Freshness (h)}}}%
      \csname LTb\endcsname%
      \put(2714,130){\makebox(0,0){\strut{}Batch number}}%
    }%
    \gplgaddtomacro\gplfronttext{%
      \csname LTb\endcsname%
      \put(3685,2168){\makebox(0,0)[r]{\strut{}\FO}}%
      \csname LTb\endcsname%
      \put(3685,1982){\makebox(0,0)[r]{\strut{}\INT}}%
      \csname LTb\endcsname%
      \put(3685,1796){\makebox(0,0)[r]{\strut{}\TB}}%
    }%
    \gplbacktext
    \put(0,0){\includegraphics{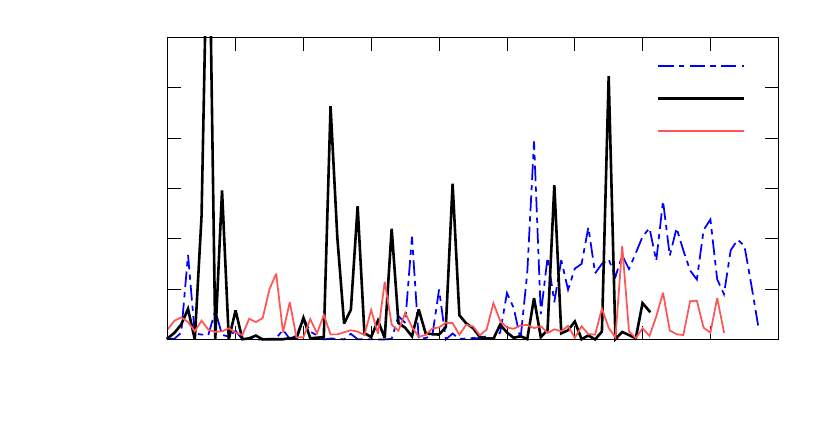}}%
    \gplfronttext
  \end{picture}%
\endgroup

%% file: input/tab-domains.tex
\small
\newcolumntype{Y}{>{\raggedright\arraybackslash}X}
  \begin{tabularx}{\textwidth}{YYYYYYYY}
\toprule
\multicolumn{4}{c}{\Ebola} & 
\multicolumn{4}{c}{\Ukraine} \\
\cmidrule(lr){1-4}
\cmidrule(lr){5-8}

\TB & \UN & \FO & \INT & \TB & \UN & \FO & \INT \\
\cmidrule(lr){1-4}
\cmidrule(lr){5-8}
       instagram.com	&            who.int	&              who.int	&       instagram.com	&         youtube.com    &          reuters.com  &              rt.com    &               rt.com  \\
          linkis.com	&        reuters.com	&              cdc.gov	&             vine.co	&    hosting-test.net    &               rt.com  &       wikipedia.org    &               ria.ru  \\
      newslocker.com	&        twitter.com	&       gnnliberia.com	&            news0.tk	&            nvua.net    &         kyivpost.com  &       wordpress.com    &        wordpress.com  \\
         nytimes.com	&      thejournal.ie	&               ria.ru	&      allnews24h .com	&       instagram .com    &            pehub.com  &              ria.ru    &         kyivpost.com  \\
  huffingtonpost .com	&         google.com	& aerzte-ohne-grenzen.at&          weeder.org	&        facebook.com    &      theguardian .com  &     theguardian .com    &      theguardian .com  \\
\bottomrule

\end{tabularx}

%% file: figures/freshness_distribution.tex
\begingroup
  \makeatletter
  \providecommand\color[2][]{%
    \GenericError{(gnuplot) \space\space\space\@spaces}{%
      Package color not loaded in conjunction with
      terminal option `colourtext'%
    }{See the gnuplot documentation for explanation.%
    }{Either use 'blacktext' in gnuplot or load the package
      color.sty in LaTeX.}%
    \renewcommand\color[2][]{}%
  }%
  \providecommand\includegraphics[2][]{%
    \GenericError{(gnuplot) \space\space\space\@spaces}{%
      Package graphicx or graphics not loaded%
    }{See the gnuplot documentation for explanation.%
    }{The gnuplot epslatex terminal needs graphicx.sty or graphics.sty.}%
    \renewcommand\includegraphics[2][]{}%
  }%
  \providecommand\rotatebox[2]{#2}%
  \@ifundefined{ifGPcolor}{%
    \newif\ifGPcolor
    \GPcolortrue
  }{}%
  \@ifundefined{ifGPblacktext}{%
    \newif\ifGPblacktext
    \GPblacktextfalse
  }{}%
  \let\gplgaddtomacro\g@addto@macro
  \gdef\gplbacktext{}%
  \gdef\gplfronttext{}%
  \makeatother
  \ifGPblacktext
    \def\colorrgb#1{}%
    \def\colorgray#1{}%
  \else
    \ifGPcolor
      \def\colorrgb#1{\color[rgb]{#1}}%
      \def\colorgray#1{\color[gray]{#1}}%
      \expandafter\def\csname LTw\endcsname{\color{white}}%
      \expandafter\def\csname LTb\endcsname{\color{black}}%
      \expandafter\def\csname LTa\endcsname{\color{black}}%
      \expandafter\def\csname LT0\endcsname{\color[rgb]{1,0,0}}%
      \expandafter\def\csname LT1\endcsname{\color[rgb]{0,1,0}}%
      \expandafter\def\csname LT2\endcsname{\color[rgb]{0,0,1}}%
      \expandafter\def\csname LT3\endcsname{\color[rgb]{1,0,1}}%
      \expandafter\def\csname LT4\endcsname{\color[rgb]{0,1,1}}%
      \expandafter\def\csname LT5\endcsname{\color[rgb]{1,1,0}}%
      \expandafter\def\csname LT6\endcsname{\color[rgb]{0,0,0}}%
      \expandafter\def\csname LT7\endcsname{\color[rgb]{1,0.3,0}}%
      \expandafter\def\csname LT8\endcsname{\color[rgb]{0.5,0.5,0.5}}%
    \else
      \def\colorrgb#1{\color{black}}%
      \def\colorgray#1{\color[gray]{#1}}%
      \expandafter\def\csname LTw\endcsname{\color{white}}%
      \expandafter\def\csname LTb\endcsname{\color{black}}%
      \expandafter\def\csname LTa\endcsname{\color{black}}%
      \expandafter\def\csname LT0\endcsname{\color{black}}%
      \expandafter\def\csname LT1\endcsname{\color{black}}%
      \expandafter\def\csname LT2\endcsname{\color{black}}%
      \expandafter\def\csname LT3\endcsname{\color{black}}%
      \expandafter\def\csname LT4\endcsname{\color{black}}%
      \expandafter\def\csname LT5\endcsname{\color{black}}%
      \expandafter\def\csname LT6\endcsname{\color{black}}%
      \expandafter\def\csname LT7\endcsname{\color{black}}%
      \expandafter\def\csname LT8\endcsname{\color{black}}%
    \fi
  \fi
  \setlength{\unitlength}{0.0500bp}%
  \begin{picture}(4780.00,2820.00)%
    \gplgaddtomacro\gplbacktext{%
      \csname LTb\endcsname%
      \put(1053,372){\makebox(0,0)[r]{\strut{} 0.0001}}%
      \csname LTb\endcsname%
      \put(1053,596){\makebox(0,0)[r]{\strut{} 0.001}}%
      \csname LTb\endcsname%
      \put(1053,821){\makebox(0,0)[r]{\strut{} 0.01}}%
      \csname LTb\endcsname%
      \put(1053,1045){\makebox(0,0)[r]{\strut{} 0.1}}%
      \csname LTb\endcsname%
      \put(1053,1269){\makebox(0,0)[r]{\strut{} 1}}%
      \csname LTb\endcsname%
      \put(1053,1494){\makebox(0,0)[r]{\strut{} 10}}%
      \csname LTb\endcsname%
      \put(1053,1718){\makebox(0,0)[r]{\strut{} 100}}%
      \csname LTb\endcsname%
      \put(1053,1942){\makebox(0,0)[r]{\strut{} 1000}}%
      \csname LTb\endcsname%
      \put(1053,2166){\makebox(0,0)[r]{\strut{} 10000}}%
      \csname LTb\endcsname%
      \put(1053,2391){\makebox(0,0)[r]{\strut{} 100000}}%
      \csname LTb\endcsname%
      \put(1053,2615){\makebox(0,0)[r]{\strut{} 1e+06}}%
      \csname LTb\endcsname%
      \put(1708,186){\makebox(0,0){\strut{}TB}}%
      \csname LTb\endcsname%
      \put(2445,186){\makebox(0,0){\strut{}UN}}%
      \csname LTb\endcsname%
      \put(3183,186){\makebox(0,0){\strut{}FO}}%
      \csname LTb\endcsname%
      \put(3920,186){\makebox(0,0){\strut{}INT}}%
      \csname LTb\endcsname%
      \put(144,1493){\rotatebox{-270}{\makebox(0,0){\strut{}Freshness ($log$ h)}}}%
    }%
    \gplgaddtomacro\gplfronttext{%
    }%
    \gplbacktext
    \put(0,0){\includegraphics{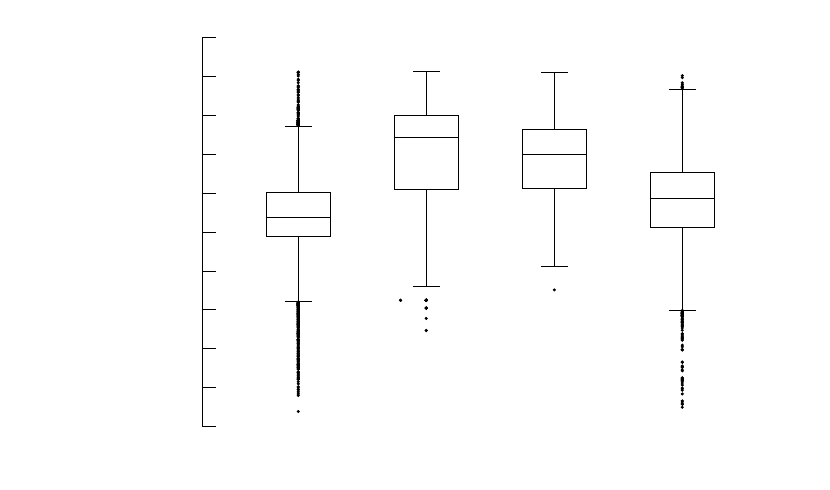}}%
    \gplfronttext
  \end{picture}%
\endgroup

%% file: figures/freshness_distribution_ukraine2.tex
\begingroup
  \makeatletter
  \providecommand\color[2][]{%
    \GenericError{(gnuplot) \space\space\space\@spaces}{%
      Package color not loaded in conjunction with
      terminal option `colourtext'%
    }{See the gnuplot documentation for explanation.%
    }{Either use 'blacktext' in gnuplot or load the package
      color.sty in LaTeX.}%
    \renewcommand\color[2][]{}%
  }%
  \providecommand\includegraphics[2][]{%
    \GenericError{(gnuplot) \space\space\space\@spaces}{%
      Package graphicx or graphics not loaded%
    }{See the gnuplot documentation for explanation.%
    }{The gnuplot epslatex terminal needs graphicx.sty or graphics.sty.}%
    \renewcommand\includegraphics[2][]{}%
  }%
  \providecommand\rotatebox[2]{#2}%
  \@ifundefined{ifGPcolor}{%
    \newif\ifGPcolor
    \GPcolortrue
  }{}%
  \@ifundefined{ifGPblacktext}{%
    \newif\ifGPblacktext
    \GPblacktextfalse
  }{}%
  \let\gplgaddtomacro\g@addto@macro
  \gdef\gplbacktext{}%
  \gdef\gplfronttext{}%
  \makeatother
  \ifGPblacktext
    \def\colorrgb#1{}%
    \def\colorgray#1{}%
  \else
    \ifGPcolor
      \def\colorrgb#1{\color[rgb]{#1}}%
      \def\colorgray#1{\color[gray]{#1}}%
      \expandafter\def\csname LTw\endcsname{\color{white}}%
      \expandafter\def\csname LTb\endcsname{\color{black}}%
      \expandafter\def\csname LTa\endcsname{\color{black}}%
      \expandafter\def\csname LT0\endcsname{\color[rgb]{1,0,0}}%
      \expandafter\def\csname LT1\endcsname{\color[rgb]{0,1,0}}%
      \expandafter\def\csname LT2\endcsname{\color[rgb]{0,0,1}}%
      \expandafter\def\csname LT3\endcsname{\color[rgb]{1,0,1}}%
      \expandafter\def\csname LT4\endcsname{\color[rgb]{0,1,1}}%
      \expandafter\def\csname LT5\endcsname{\color[rgb]{1,1,0}}%
      \expandafter\def\csname LT6\endcsname{\color[rgb]{0,0,0}}%
      \expandafter\def\csname LT7\endcsname{\color[rgb]{1,0.3,0}}%
      \expandafter\def\csname LT8\endcsname{\color[rgb]{0.5,0.5,0.5}}%
    \else
      \def\colorrgb#1{\color{black}}%
      \def\colorgray#1{\color[gray]{#1}}%
      \expandafter\def\csname LTw\endcsname{\color{white}}%
      \expandafter\def\csname LTb\endcsname{\color{black}}%
      \expandafter\def\csname LTa\endcsname{\color{black}}%
      \expandafter\def\csname LT0\endcsname{\color{black}}%
      \expandafter\def\csname LT1\endcsname{\color{black}}%
      \expandafter\def\csname LT2\endcsname{\color{black}}%
      \expandafter\def\csname LT3\endcsname{\color{black}}%
      \expandafter\def\csname LT4\endcsname{\color{black}}%
      \expandafter\def\csname LT5\endcsname{\color{black}}%
      \expandafter\def\csname LT6\endcsname{\color{black}}%
      \expandafter\def\csname LT7\endcsname{\color{black}}%
      \expandafter\def\csname LT8\endcsname{\color{black}}%
    \fi
  \fi
  \setlength{\unitlength}{0.0500bp}%
  \begin{picture}(4780.00,2820.00)%
    \gplgaddtomacro\gplbacktext{%
      \csname LTb\endcsname%
      \put(1053,372){\makebox(0,0)[r]{\strut{} 0.0001}}%
      \csname LTb\endcsname%
      \put(1053,596){\makebox(0,0)[r]{\strut{} 0.001}}%
      \csname LTb\endcsname%
      \put(1053,821){\makebox(0,0)[r]{\strut{} 0.01}}%
      \csname LTb\endcsname%
      \put(1053,1045){\makebox(0,0)[r]{\strut{} 0.1}}%
      \csname LTb\endcsname%
      \put(1053,1269){\makebox(0,0)[r]{\strut{} 1}}%
      \csname LTb\endcsname%
      \put(1053,1494){\makebox(0,0)[r]{\strut{} 10}}%
      \csname LTb\endcsname%
      \put(1053,1718){\makebox(0,0)[r]{\strut{} 100}}%
      \csname LTb\endcsname%
      \put(1053,1942){\makebox(0,0)[r]{\strut{} 1000}}%
      \csname LTb\endcsname%
      \put(1053,2166){\makebox(0,0)[r]{\strut{} 10000}}%
      \csname LTb\endcsname%
      \put(1053,2391){\makebox(0,0)[r]{\strut{} 100000}}%
      \csname LTb\endcsname%
      \put(1053,2615){\makebox(0,0)[r]{\strut{} 1e+06}}%
      \csname LTb\endcsname%
      \put(1708,186){\makebox(0,0){\strut{}TB}}%
      \csname LTb\endcsname%
      \put(2445,186){\makebox(0,0){\strut{}UN}}%
      \csname LTb\endcsname%
      \put(3183,186){\makebox(0,0){\strut{}FO}}%
      \csname LTb\endcsname%
      \put(3920,186){\makebox(0,0){\strut{}INT}}%
      \csname LTb\endcsname%
      \put(144,1493){\rotatebox{-270}{\makebox(0,0){\strut{}Freshness ($log$ h)}}}%
    }%
    \gplgaddtomacro\gplfronttext{%
    }%
    \gplbacktext
    \put(0,0){\includegraphics{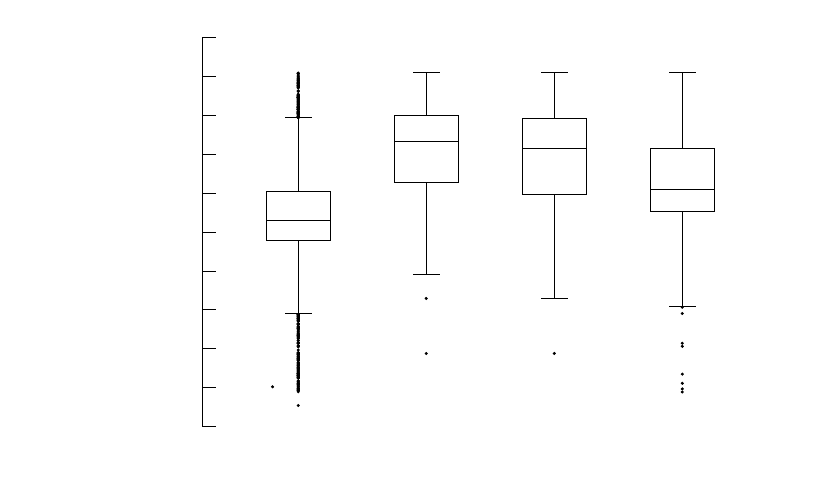}}%
    \gplfronttext
  \end{picture}%
\endgroup

%% file: input/related.tex

Web crawlers are typically developed in the context of Web search applications.
General considerations for these crawlers are described by \citep[][chap. 20]{manning2008}, 
with more recent developments summarized in \cite{olston2010}.
For Web archiving, the web crawler Heritrix by the Internet Archive \cite{mohr04heritrix} is commonly used.
Standard crawling methods aim to capture as much of the Web as possible.
In contrast, \emph{focused crawling} \cite{chakrabarti99:focus} aims to only crawl pages that are related to a 
specific topic.
Focused crawlers \citep[e.g.~][]{Aggarwal:2001:ICW:371920.371955,Pant2005}
learn a representation of the topic from the set of initial pages (\emph{seed URLs}) and follow links only if the 
containing page matches that representation.
Extensions of this model use ontologies to incorporate semantic knowledge into the matching 
process \cite{Ehrig2003,Dong13}, `tunnel' between disjoint page 
clusters \cite{bergmark_focused_2002,jialun_qin_building_2004} or learn navigation 
structures necessary to find relevant pages~\cite{Diligenti2000,jiang_focus_2013}.
In the recently proposed time-aware focused crawling \cite{PereiraMCM14} time is used as a primary focusing 
criteria. Here the crawler is guided to follow links that are related to a target time,
but the topical relevance is not considered. 
In summary, existing solutions to focused Web crawling consider relevance and time dimensions
in isolation and do not address the problem of jointly finding relevant as well as fresh content.

The Social Web provides an important source of data for Web Science researchers.
Many services such as Twitter, 
Youtube or 
Flickr
provide access to structured information about users, user networks and created content through their APIs 
and are therefore attractive to researchers.
Data collection from these services is not supported by standard Web crawlers.
Usually it is conducted in an ad-hoc manner, although some structured approaches targeting specific aspects
exist \cite{Boanjak:2012:TDF:2187980.2188266, psallidas2013}.
For example, \cite{Boanjak:2012:TDF:2187980.2188266} collects Twitter data 
from particular user communities, \cite{psallidas2013} proposes a cross Social Media crawler,
whereas \cite{Aiello2013} addresses topic detection in Twitter streams.
These studies typically focus on crawling and analyzing data from specific Social Networks, whereas our work addresses the problem of integrated collection of interlinked Web and Social Web data.

%

The potential relevance of Tweets for Web archive creation has been explored \cite{Yang:2012}.
In the ARCOMEM project \cite{RisseFI:2014} first approaches have been investigated to 
implement a social and semantic driven selection model for Web and Social Web content.
The results show that combination of social and semantic information can lead to focused Web archives.
However, their system has separate Web and Social Media crawlers which causes 
a drift of focus between the subsystems.
%
%
In contrast, iCrawl is a fully integrated crawler to seamlessly collect interlinked 
Web and Social Web content and guide the focused Web crawler using Social Media.
  



Data freshness is a data quality dimension that has various application-dependent 
definitions and metrics \cite{Bouzeghoub:2004}. The aspects of freshness include e.g. \textit{currency}, i.e. the time interval 
between the data extraction and its delivery to the user, and \textit{timeliness}, i.e.\@ the actual age of the data. 
%
Current search engines use the freshness of documents as part of their scoring algorithms.
Here freshness is estimated based on the crawling history associated with the documents (such as 
the inception date, i.e.\@ the date the document has been discovered and indexed by the search engine or
the date it first appeared in the search results) \cite{dong2011Google}.
Even though content freshness is one of the most important requirements of the iCrawl users, 
they cannot always rely on the crawl history to estimate freshness of pages. Instead, iCrawl relies on Social Media 
to provide entry points to the fresh content and uses content-based freshness estimates for evaluation.




%% file: input/conclusion.tex

In this paper we addressed the problem of collection of fresh and relevant Web and Social Web 
content for current events and topics of interest. 
To achieve thematically coherent and fresh Web collections, we proposed 
the novel paradigm of integrated crawling that exploits Social Media streams and interlinking 
between the Web and Social Web content to continuously guide
a focused crawler towards fresh and relevant content.
We presented iCrawl (available online at \url{http://icrawl.l3s.uni-hannover.de}), an open source integrated 
focused crawler that 
seamlessly connects focused Web crawling and Twitter API query 
in one system and enables scalable and efficient collection of interlinked fresh Web and Social Web content
on a topic of interest.
We confirmed, that Twitter can be effectively used as a source of fresh content.
%
Our experiments with real-world datasets collected by iCrawl demonstrate that 
the integrated crawler takes the advantage of both precise focused crawling and 
continuous input of fresh Social Media streams and automatically adapts
its behaviour towards the most promising information source.


\subsection*{Acknowledgements}

This work was partially funded by the ERC under ALEXANDRIA
(ERC 339233) and BMBF (Project ``GlycoRec''). 

%% file: icrawl_jcdl15.bbl
\begin{thebibliography}{26}
\providecommand{\natexlab}[1]{#1}
\providecommand{\url}[1]{\texttt{#1}}
\expandafter\ifx\csname urlstyle\endcsname\relax
  \providecommand{\doi}[1]{doi: #1}\else
  \providecommand{\doi}{doi: \begingroup \urlstyle{rm}\Url}\fi

\bibitem[nut()]{nutch}
{Apache Nutch:} {Highly extensible, highly scalable Web crawler}.
\newblock Available online: \url{http://nutch.apache.org/} (accessed on 23
  October 2014).

\bibitem[Abiteboul et~al.(2003)Abiteboul, Preda, and Cobena]{Abiteboul2003}
S.~Abiteboul, M.~Preda, and G.~Cobena.
\newblock Adaptive on-line page importance computation.
\newblock In \emph{World Wide Web Conference}, WWW '03, 2003.
\newblock \doi{10.1145/775152.775192}.

\bibitem[Aggarwal et~al.(2001)Aggarwal, Al-Garawi, and
  Yu]{Aggarwal:2001:ICW:371920.371955}
C.~Aggarwal, F.~Al-Garawi, and P.~S. Yu.
\newblock Intelligent crawling on the world wide web with arbitrary predicates.
\newblock In \emph{World Wide Web Conference}, pages 96--105, 2001.
\newblock \doi{10.1145/371920.371955}.

\bibitem[Aiello et~al.(2013)Aiello, Petkos, Martin, Corney, Papadopoulos,
  Skraba, Goker, Kompatsiaris, and Jaimes]{Aiello2013}
L.~Aiello, G.~Petkos, C.~Martin, D.~Corney, S.~Papadopoulos, R.~Skraba,
  A.~Goker, I.~Kompatsiaris, and A.~Jaimes.
\newblock Sensing trending topics in twitter.
\newblock \emph{IEEE Transactions on Multimedia}, 15\penalty0 (6):\penalty0
  1268--1282, Oct 2013.
\newblock \doi{10.1109/TMM.2013.2265080}.

\bibitem[Bergmark et~al.(2002)Bergmark, Lagoze, and
  Sbityakov]{bergmark_focused_2002}
D.~Bergmark, C.~Lagoze, and A.~Sbityakov.
\newblock Focused crawls, tunneling, and digital libraries.
\newblock In \emph{Research and Advanced Technology for Digital Libraries}.
  Springer, 2002.

\bibitem[Boanjak et~al.(2012)Boanjak, Oliveira, Martins, Mendes~Rodrigues, and
  Sarmento]{Boanjak:2012:TDF:2187980.2188266}
M.~Boanjak, E.~Oliveira, J.~Martins, E.~Mendes~Rodrigues, and L.~Sarmento.
\newblock Twitterecho: A distributed focused crawler to support open research
  with twitter data.
\newblock In \emph{World Wide Web Conference Companion}, pages 1233--1240,
  2012.
\newblock \doi{10.1145/2187980.2188266}.

\bibitem[Bouzeghoub(2004)]{Bouzeghoub:2004}
M.~Bouzeghoub.
\newblock A framework for analysis of data freshness.
\newblock In \emph{Proceedings of the Workshop on Information Quality in
  Information Systems}, IQIS '04, pages 59--67, 2004.
\newblock \doi{10.1145/1012453.1012464}.

\bibitem[Chakrabarti et~al.(1999)Chakrabarti, van~den Berg, and
  Dom]{chakrabarti99:focus}
S.~Chakrabarti, M.~van~den Berg, and B.~Dom.
\newblock Focused crawling: a new approach to topic-specific web resource
  discovery.
\newblock \emph{Computer Networks}, 31\penalty0 (11-16):\penalty0 1623--1640,
  1999.
\newblock \doi{10.1016/S1389-1286(99)00052-3}.

\bibitem[Diligenti et~al.(2000)Diligenti, Coetzee, Lawrence, Giles, and
  Gori]{Diligenti2000}
M.~Diligenti, F.~Coetzee, S.~Lawrence, C.~L. Giles, and M.~Gori.
\newblock Focused crawling using context graphs.
\newblock In \emph{Conference on Very Large Data Bases}, pages 527--534, 2000.

\bibitem[Dong et~al.(2011)Dong, Chang, Zhang, Zheng, Mishne, Bai, Buchner,
  Liao, Ji, Leung, et~al.]{dong2011Google}
A.~Dong, Y.~Chang, R.~Zhang, Z.~Zheng, G.~Mishne, J.~Bai, K.~Buchner, C.~Liao,
  S.~Ji, G.~Leung, et~al.
\newblock Incorporating recency in network search using machine learning,
  Apr.~21 2011.
\newblock US Patent App. 12/579,855.

\bibitem[Dong and Hussain(2013)]{Dong13}
H.~Dong and F.~K. Hussain.
\newblock {SOF}: a semi-supervised ontology-learning-based focused crawler.
\newblock \emph{Concurrency and Computation: Practice and Experience},
  25\penalty0 (12):\penalty0 1755--1770, 2013.

\bibitem[Ehrig and Maedche(2003)]{Ehrig2003}
M.~Ehrig and A.~Maedche.
\newblock Ontology-focused crawling of web documents.
\newblock In \emph{ACM Symposium on Applied Computing}, pages 1174--1178, 2003.
\newblock \doi{10.1145/952532.952761}.

\bibitem[Gossen et~al.(2015)Gossen, Demidova, and
  Risse]{gerhardgossen2015icrawl}
G.~Gossen, E.~Demidova, and T.~Risse.
\newblock The {iCrawl} {Wizard} -- supporting interactive focused crawl
  specification.
\newblock In \emph{Proceedings of the European Conference on Information
  Retrieval (ECIR) 2015}, 2015.

\bibitem[Jiang et~al.(2013)Jiang, Song, Yu, and Lin]{jiang_focus_2013}
J.~Jiang, X.~Song, N.~Yu, and C.-Y. Lin.
\newblock Focus: Learning to crawl web forums.
\newblock \emph{IEEE Transactions on Knowledge and Data Engineering},
  25\penalty0 (6):\penalty0 1293--1306, June 2013.
\newblock \doi{10.1109/TKDE.2012.56}.

\bibitem[Manning et~al.(2008)Manning, Raghavan, and Sch\"{u}tze]{manning2008}
C.~D. Manning, P.~Raghavan, and H.~Sch\"{u}tze.
\newblock \emph{Introduction to Information Retrieval}.
\newblock Cambridge University Press, New York, NY, USA, 2008.
\newblock ISBN 0521865719.

\bibitem[Mohr et~al.(2004)Mohr, Kimpton, Stack, and Ranitovic]{mohr04heritrix}
G.~Mohr, M.~Kimpton, M.~Stack, and I.~Ranitovic.
\newblock Introduction to {Heritrix}, an archival quality web crawler.
\newblock In \emph{{4th International Web Archiving Workshop (IWAW04)}}, 2004.

\bibitem[Olston and Najork(2010)]{olston2010}
C.~Olston and M.~Najork.
\newblock Web crawling.
\newblock \emph{Foundations and Trends in Information Retrieval}, 4\penalty0
  (3):\penalty0 175--246, 2010.
\newblock \doi{10.1561/1500000017}.

\bibitem[Pant and Srinivasan(2005)]{Pant2005}
G.~Pant and P.~Srinivasan.
\newblock Learning to crawl: Comparing classification schemes.
\newblock \emph{ACM Transactions on Information Systems}, 23\penalty0
  (4):\penalty0 430--462, Oct. 2005.
\newblock \doi{10.1145/1095872.1095875}.

\bibitem[Pereira et~al.(2014)Pereira, Macedo, Craveiro, and
  Madeira]{PereiraMCM14}
P.~Pereira, J.~Macedo, O.~Craveiro, and H.~Madeira.
\newblock Time-aware focused web crawling.
\newblock In \emph{European Conference on {IR} Research, {ECIR} 2014}, pages
  534--539, 2014.
\newblock \doi{10.1007/978-3-319-06028-6_53}.

\bibitem[Psallidas et~al.(2013)Psallidas, Ntoulas, and Delis]{psallidas2013}
F.~Psallidas, A.~Ntoulas, and A.~Delis.
\newblock Soc web: Efficient monitoring of social network activities.
\newblock In \emph{Web Information Systems Engineering 2013}, pages 118--136.
  Springer, 2013.
\newblock \doi{10.1007/978-3-642-41154-0_9}.

\bibitem[Qin et~al.(2004)Qin, Zhou, and Chau]{jialun_qin_building_2004}
J.~Qin, Y.~Zhou, and M.~Chau.
\newblock Building domain-specific web collections for scientific digital
  libraries.
\newblock In \emph{Joint {ACM/IEEE} Conference on Digital Libraries, 2004},
  pages 135--141, June 2004.
\newblock \doi{10.1109/JCDL.2004.1336110}.

\bibitem[Risse et~al.(2014{\natexlab{a}})Risse, Demidova, Dietze, Peters,
  Papailiou, Doka, Stavrakas, Plachouras, Senellart, Carpentier, Mantrach,
  Cautis, Siehndel, and Spiliotopoulos]{RisseFI:2014}
T.~Risse, E.~Demidova, S.~Dietze, W.~Peters, N.~Papailiou, K.~Doka,
  Y.~Stavrakas, V.~Plachouras, P.~Senellart, F.~Carpentier, A.~Mantrach,
  B.~Cautis, P.~Siehndel, and D.~Spiliotopoulos.
\newblock The {ARCOMEM} architecture for social- and semantic-driven web
  archiving.
\newblock \emph{Future Internet}, 6\penalty0 (4):\penalty0 688--716,
  2014{\natexlab{a}}.
\newblock ISSN 1999-5903.

\bibitem[Risse et~al.(2014{\natexlab{b}})Risse, Demidova, and
  Gossen]{icrawlRequirements}
T.~Risse, E.~Demidova, and G.~Gossen.
\newblock What do you want to collect from the web?
\newblock In \emph{Proc. of the Building Web Observatories Workshop (BWOW)
  2014}, 2014{\natexlab{b}}.

\bibitem[SalahEldeen and Nelson(2012)]{salaheldeen2012}
H.~M. SalahEldeen and M.~L. Nelson.
\newblock Losing my revolution: How many resources shared on social media have
  been lost?
\newblock In \emph{Theory and Practice of Digital Libraries}, pages 125--137.
  Springer, 2012.
\newblock \doi{10.1007/978-3-642-33290-6_14}.

\bibitem[Tannier(2014)]{Tannier14}
X.~Tannier.
\newblock Extracting news web page creation time with {DCTFinder}.
\newblock In \emph{Conference on Language Resources and Evaluation
  (LREC-2014)}, pages 2037--2042, 2014.

\bibitem[Yang et~al.(2012)Yang, Chitturi, Wilson, Magdy, and Fox]{Yang:2012}
S.~Yang, K.~Chitturi, G.~Wilson, M.~Magdy, and E.~A. Fox.
\newblock A study of automation from seed {URL} generation to focused web
  archive development: The {CTRnet} context.
\newblock In \emph{Proceedings of the 12th ACM/IEEE-CS Joint Conference on
  Digital Libraries}, JCDL '12, pages 341--342, 2012.
\newblock \doi{10.1145/2232817.2232881}.

\end{thebibliography}
